\documentclass[%
reprint,
amsmath,amssymb,
aps,
prx,
onecolumn,
longbibliography
]{revtex4-2}

\usepackage{graphicx}
\usepackage{epsfig}
\usepackage{dcolumn}
\usepackage{bm}
\usepackage{epstopdf}
\usepackage{multirow}
\usepackage{amsmath}
\usepackage{booktabs}
\usepackage{color}
\usepackage{textcomp, gensymb}
\usepackage{kantlipsum}  
\usepackage[colorlinks=true, allcolors=blue]{hyperref}
\usepackage{ulem}
\usepackage{braket}
\usepackage{physics}
\usepackage{url}
\usepackage[utf8]{inputenc}
\usepackage[T1]{fontenc}
\usepackage{mathptmx}
\usepackage{lineno}
\usepackage{tcolorbox}
\usepackage{hyperref}
\usepackage{lipsum}
\usepackage{tabularray}
\usepackage[skip=8pt]{parskip}
\definecolor{green}{rgb}{.2,.6,.2}
\definecolor{brickred}{rgb}{0.8, 0.25, 0.33}
\definecolor{brightcerulean}{rgb}{0.11, 0.67, 0.84}
\definecolor{maroon}{rgb}{0.788, 0.0, 0.086}
\definecolor{ao}{rgb}{0.0, 0.5, 0.0}

\newcommand{\pd}[1]{\partial_{#1}}

\newcommand{\stress}[1]{\textit{#1}}

\newcommand{\intra}{\textit{v}} 
\newcommand{\inter}{\textit{w}} 
\newcommand{\sshintra}{\textit{m}} 
\newcommand{\sshinter}{\textit{t}} 
\renewcommand{\i}{\ensuremath i} 
\newcommand{\windnum}{\vartheta} 

\newcommand{\fig}{figure}
\newcommand{\figs}{figures}
\newcommand{\tab}{table}
\newcommand{\app}{appendix}

\newcommand{\seclab}{section}
\newcommand{\seclabs}{sections}
\newcommand{\eq}{equation}
\newcommand{\eqs}{equations}
\newcommand{\reflab}{Ref.}
\newcommand{\reflabs}{Refs.}

\newcommand{\mbc}[1]{\hat{c}_{#1}} 
\newcommand{\mbcd}[1]{\hat{c}^\dagger_{#1}} 

\begin{document}

\title{Characterizing out-of-distribution generalization of neural networks: \\application to the disordered Su-Schrieffer-Heeger model}

\author{Kacper~Cybi\'nski$^1$}
\author{Marcin~Płodzień$^2$}
\author{Micha{\l}~Tomza$^1$}
\author{Maciej~Lewenstein$^{2,3}$}
\author{Alexandre~Dauphin$^{2,4}$}
\author{Anna~Dawid$^5$}

\affiliation{
$^1$Faculty of Physics, University of Warsaw, Pasteura 5, 02-093 Warsaw, Poland
}
\affiliation{
$^2$ICFO - Institut de Ci\`encies Fot\`oniques, The Barcelona Institute of Science and Technology, Av. Carl Friedrich Gauss 3, 08860 Castelldefels (Barcelona), Spain
}
\affiliation{
$^3$ICREA, Pg.~Llu\'{i}s Campanys 23, 08010 Barcelona, Spain
}
\affiliation{
$^4$PASQAL SAS, 2 av. Augustin Fresnel Palaiseau, 91120, France
}
\affiliation{
$^5$Center for Computational Quantum Physics, Flatiron Institute, 162 Fifth Avenue, New York, NY 10010, USA
}
\date{\today}

\begin{abstract}
Machine learning (ML) is a promising tool for the detection of phases of matter. However, ML models are also known for their black-box construction, which hinders understanding of what they learn from the data and makes their application to novel data risky. Moreover, the central challenge of ML is to ensure its good generalization abilities, i.e., good performance on data outside the training set. Here, we show how the informed use of an interpretability method called class activation mapping (CAM), and the analysis of the latent representation of the data with the principal component analysis (PCA) can increase trust in predictions of a neural network (NN) trained to classify quantum phases. In particular, we show that we can ensure better out-of-distribution generalization in the complex classification problem by choosing such an NN that, in the simplified version of the problem, learns a known characteristic of the phase. We show this on an example of the topological Su–Schrieffer–Heeger (SSH) model with and without disorder, which turned out to be surprisingly challenging for NNs trained in a supervised way. This work is an example of how the systematic use of interpretability methods can improve the performance of NNs in scientific problems.
\end{abstract}

\maketitle

\section{Introduction and motivation}

Machine learning (ML) promises a revolution in science, similar to the current revolution in industry~\cite{dawid2023introduction}. In quantum physics, neural networks (NNs) serve as a flexible and promising representation of quantum states~\cite{Hermann2023NQS, medvidovic2024NQS, lange2024architectures} and a booster for quantum technologies~\cite{dawid2023modern, Krenn23MLforQT, Mafu2024advances}, e.g., as entanglement classifiers~\cite{Lu2018separability, Sanavio2023entanglement}. Neural networks are especially promising in the detection of phases of matter and have been used for classical~\cite{Carrasquilla17NatPhys, li_applications_2018, schafer_vector_2019, bachtis2020mapping, Liu2021, richter2023learning}, quantum \cite{van_nieuwenburg_learning_2017, wetzel_unsupervised_2017, liu_discriminative_2018, chng_unsupervised_2018, huembeli_automated_2019, kottmann_unsupervised_2020, kottmann2021unsupervised, arnold_interpretable_2021, patel2022unsupervised, broecker_machine_2017, theveniaut_neural_2019, dong_machine_2019, blucher_towards_2020}, and topological~\cite{zhang_machine_2018, tsai2020deep, Baireuther23identifying, huembeli_identifying_2018, rodriguez-nieva_identifying_2019, greplova_unsupervised_2020, Balabanov2021, yu2023unsupervised, teng2023clustering} phase transitions with supervised \cite{Carrasquilla17NatPhys, broecker_machine_2017, theveniaut_neural_2019, dong_machine_2019, blucher_towards_2020, li_applications_2018, zhang_machine_2018, tsai2020deep, Baireuther23identifying} and unsupervised \cite{huembeli_identifying_2018, bachtis2020mapping, schafer_vector_2019, van_nieuwenburg_learning_2017, wetzel_unsupervised_2017, liu_discriminative_2018, chng_unsupervised_2018, greplova_unsupervised_2020, huembeli_automated_2019, rodriguez-nieva_identifying_2019, kottmann_unsupervised_2020, kottmann2021unsupervised, arnold_interpretable_2021, Balabanov2021, Liu2021, patel2022unsupervised, yu2023unsupervised, teng2023clustering, richter2023learning} approaches as well as for experimental data \cite{Rem19, Khatami20, Kaming2021, miles2023machine, link2023machine}.
Other examples include ML approaches relying on dimensionality reduction \cite{Wang16, che2020topological, long2020unsupervised, yang2021visualizing}, kernel methods \cite{Vargas18b, Greitemann19, Greitemann2021}, topological data analysis \cite{leykam2023topological, scheurer_unsupervised_2020, cole_quantitative_2021, park2022unsupervised}, and quantum NNs \cite{schuld2019quantum, caro2022generalization, liu2023model}.

However, before NNs join standard toolboxes for the analysis of phases of matter, they need to become more interpretable (so we understand what they learn) and reliable (so we can trust their predictions). Note that interpretability is a stronger condition than reliability (if we understand exactly how an NN makes its predictions, we usually can trust it). There are extensive efforts to make automated approaches more interpretable~\cite{wetzel_unsupervised_2017, Dawid20NJP, wetzel_discovering_2020, arnold_interpretable_2021, cole_quantitative_2021, Dawid2021Hessian, Balabanov2021, arnold2022replacing, wetzel2024closedform}, which should ultimately lead to learning phases of matter and assisting physicists in understanding the learned order parameters~\cite{Wetzel17b, Greitemann19probing, Liu19interpretable, Miles2021CCNN, miles2023machine}. However, the community focuses predominantly on a special representation of data, that is, spin configurations. Therefore, the question of the interpretability and reliability of NNs applied to different quantum data remains wide open.

Reliable NNs are expected to generalize robustly to new scenarios \cite{tran2022plex}. It is especially challenging in the case of the out-of-distribution (OOD) generalization when the test data come from a different distribution than the training data. Naturally, no OOD generalization should be expected for unrelated training and test distributions. However, a robust model also performs well under a limited distribution shift. Such an OOD generalization of a network can be checked when we have enough information on the test distribution; for example, in a supervised scenario, we have access to labels of some OOD test data. When we do not have such labels, our trust in the OOD generalization of an NN has to be limited, especially in the presence of spurious correlations in the training data.

\begin{figure}[t]
\begin{center}
\includegraphics[width=0.95\columnwidth]{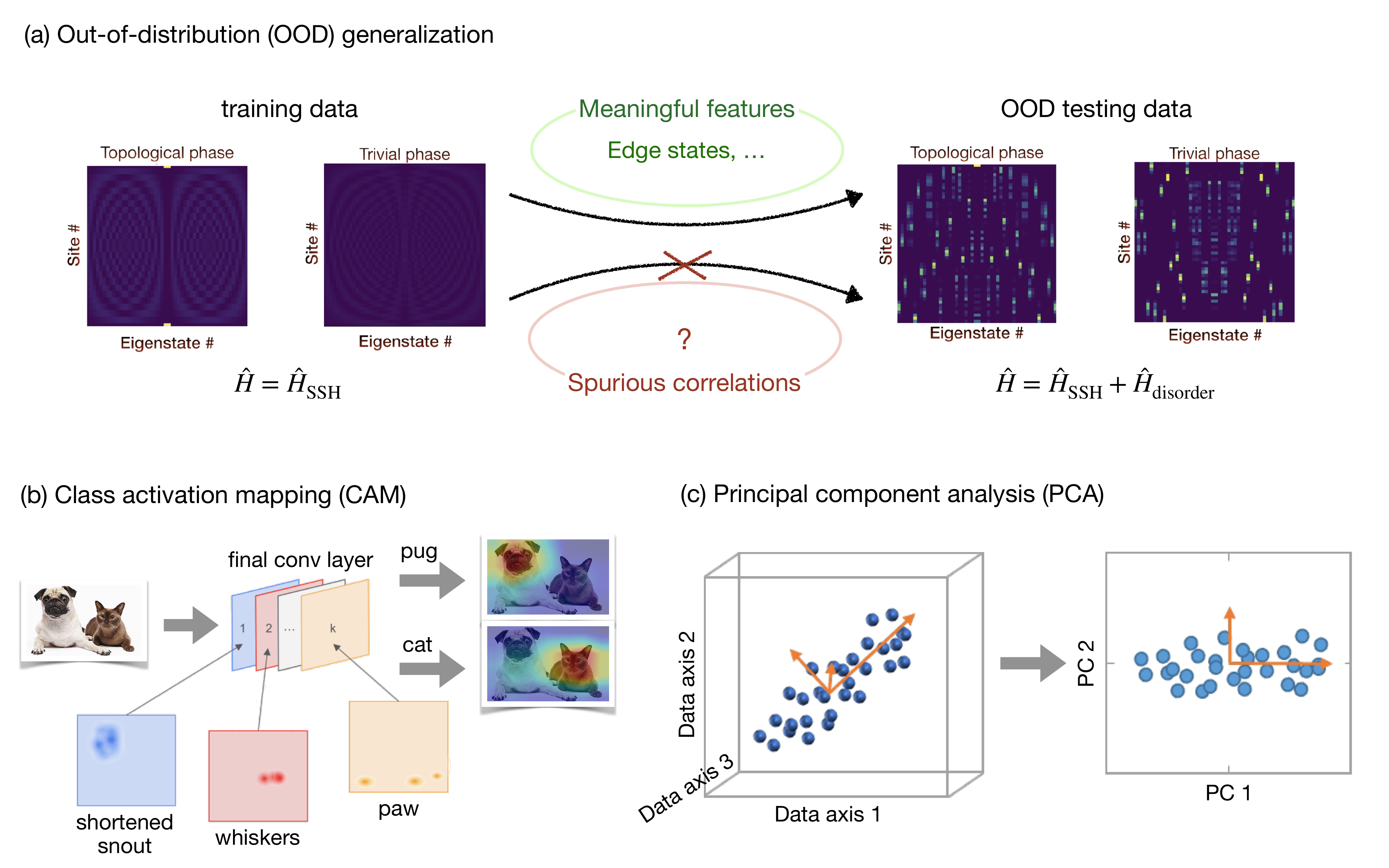}
\end{center}
\caption{\textbf{Unsupervised study of out-of-distribution generalization with class activation mapping and principal component analysis.} (a) We train multiple randomly initialized CNNs to predict phases in the SSH model without the disorder (simple regime) and study their OOD generalization to phases in the SSH model with disorder (complex regime). Only a handful of trained CNNs learn meaningful features (instead of spurious correlations) that allow for successful OOD generalization. (b) We analyze what data features are the most important for predictions of well- and poorly-generalizing CNNs with class activation mapping (CAM) -- an interpretability technique producing heatmaps that after overlaying on input samples indicate the most important data elements for the prediction. (c) With principal component analysis (PCA), we study the similarity of the representation learned by poorly- and well-generalizing CNNs for the data from the simple and complex regime. Panels (b) and (c) are adapted from \reflabs~\cite{fischer_CAM_2022, Dawid2022PhD}, respectively.}
\label{fig:intro}
\end{figure}

In this work, we show how we can increase trust in the OOD generalization of an NN in the absence of labeled OOD data. To this end, we study explanations of NN predictions and the representation of data learned by an NN and discuss patterns that correlate well with the network robustness. We perform this analysis on an example of data coming from a prototypical topological Hamiltonian, that is, the Su-Schrieffer-Heeger (SSH) model. The training data come from the standard SSH model, while the OOD test data are from the SSH model with the disorder, as presented schematically in \fig~\ref{fig:intro}(a).

Previous work~\cite{huembeli_identifying_2018} showed that standard convolutional neural networks (CNNs) trained on the data from the SSH model struggle to generalize under the disorder. The authors solved this problem by using a domain adversarial neural network. Here, we do not aim to improve this solution. Instead, we want to understand the reason for the failures of standard CNNs and propose tools that can inform the user whether to expect an OOD generalization from a trained CNN or not without labeled data.

This paper is structured as follows. We start by describing the data set and the learning task of a CNN in \seclabs~\ref{sec:Hamiltonian}-\ref{sec:task}. In \seclab~\ref{sec:CAM}, we discuss an interpretability technique called class activation mapping (CAM) \cite{Zhou16CAM}, which provides explanations of CNN predictions as sketched in \fig~\ref{fig:intro}(b). To study the data representation learned by a CNN, we need a dimensionality reduction technique such as principal component analysis (PCA), presented schematically in \fig~\ref{fig:intro}(c), which we describe in \seclab~\ref{sec:PCA_UMAP}. We present and discuss our results in \seclab~\ref{sec:results_discussion} and conclude in \seclab~\ref{sec:conclusion}.

\section{Methods}

\subsection{The model Hamiltonian}\label{sec:Hamiltonian}

The model proposed by Su, Schrieffer, and Heeger (SSH) \cite{asboth_short_2016} describes spinless fermions on a one-dimensional (1D) chain in a tight-binding approximation, with staggered nearest-neighbor tunneling amplitudes. It is described by the Hamiltonian
\begin{equation}\label{eq:SSH_clean_ham}
    \hat{H}_0 = \intra \sum_{n=1}^N \mbcd{n} \hat{\sigma}_x \mbc{n} + \inter \sum_{n=1}^{N-1} \qty(\frac{1}{2}\mbcd{n}\qty(\hat{\sigma}_x + \i \hat{\sigma}_y)\mbc{n+1} +\text{h.c.}).
\end{equation}
The SSH Hamiltonian can have alternative definitions depending on the conventions~\cite{mondragon-shem_topological_2014, meier_observation_2018, le_topological_2020} (see \app~\ref{app:formulations} for more details). We consider a chain of $N$ unit cells. Within each unit cell, there are two sites that belong to sublattice $A$ and $B$, respectively. The operator $\mbc{i}^{(\dagger)} = \qty(\mbc{A, i}^{(\dagger)}, \mbc{B, i}^{(\dagger)})$ denotes the annihilation(creation) operators on the respective sublattices in the system. $\sigma_{i}$ denotes the Pauli matrices. The tunneling amplitudes are different for intercell tunneling (between neighboring cells) -- $\inter$ and for intracell tunneling (within a single unit cell) -- $\intra$. For the remainder of this work, we fix the intercell tunneling to $\inter = 1$.

Such a system exhibits two distinct phases: topological and trivial. They are characterized by a different value of a \textit{topological invariant}, called the winding number, which arises from the chiral symmetry present in the system. 
For periodic boundary conditions (PBC), the invariant is \textit{winding number} $\windnum$. For open boundary conditions (OBC), the topological phase is characterized by the real space winding number and by the presence of zero-energy eigenstates, corresponding to the half-filling of the system~\cite{meier_observation_2018}. These states are exponentially localized at the edges of the system. For this reason, they are referred to as \textit{edge states} or \textit{edge modes}. The number of such zero-energy edge states on one edge is equal the winding number of the bulk. It can be shown by bulk-boundary correspondence~\cite{asboth_short_2016}, that both formulations are equivalent; therefore, in this work we use both approaches. 

The interplay between the topological invariants of the system and the presence of the disorder can be studied with the following Hamiltonian~\cite{mondragon-shem_topological_2014},
\begin{equation} \label{eq:ssh_disordered_ham}
    \hat{H}_1 =\sum_{n=1}^{N-1} t_n\qty(\frac{1}{2} \hat{c}_n^{\dagger}\left(\hat{\sigma}_x+\i \hat{\sigma}_y\right) \hat{c}_{n+1}+\text { h.c. }) +
    \sum_{n=1}^N m_n \hat{c}_n^{\dagger} \hat{\sigma}_x \hat{c}_n.
\end{equation}
Here, the tunneling amplitudes $t_n$ and $m_n$ vary from site to site. They include the added disorder in the following way
\begin{equation}
    t_n = \inter + 2W \omega_n \qc m_n = \intra + W \omega'_n.
\end{equation}
The parameter $W$ is the strength of the disorder, and $\omega_n$ is a random variable drawn from the uniform distribution on the interval $[-0.5, 0.5]$. The topological invariants, to some extent, are robust to disorder. However, for a strong enough value of the disorder, they are likely to change. Therefore, to account for disorder in winding number calculation, we use a handy approximation based on infinite system winding number calculation~\cite{fraxanet_topological_2021} (see \app~\ref{app:windnum} for details). The relationship between winding number $\windnum$ and disorder strength $W/\inter$ for systems of the symmetry class such as \eq~\eqref{eq:ssh_disordered_ham} has been thoroughly studied in \reflab~\cite{mondragon-shem_topological_2014}, where they proposed a phase diagram for this scenario. The phase diagram we calculate using the infinite system winding number approximation is in good agreement with their results.

\subsection{Learning task and data set}\label{sec:task}

In this work, we task a convolutional neural network (CNN) with learning a mapping between a full set of eigenstates of the SSH Hamiltonian from \eq~\eqref{eq:SSH_clean_ham} and the corresponding winding number. We describe the CNN architecture and training hyperparameters in detail in \app~\ref{app:architectures}. To obtain the input data, we compute the eigenstates of \eq~\eqref{eq:SSH_clean_ham} using exact diagonalization~\cite{zhang_exact_2010} with OBC. The corresponding winding numbers are computed by taking the same Hamiltonian with the imposed PBC. This is an approximate calculation, which translates into calculating the winding number of an infinite system~\cite{fraxanet_topological_2021}.

The input data are matrices whose columns are squared moduli of the eigenstates of the system, sorted by energy. We present exemplary input data samples in \fig~\ref{fig:intro}(a). The labels are corresponding winding numbers. Because of the presence of chiral symmetry, the energy spectrum is symmetric, and the zero-energy edge states correspond to the two middle columns of an input data sample. The amplitude of the edge states peaks at the edges and decays exponentially into the bulk; therefore, we expect the edge pixels of the middle columns to be the most indicative of these states in the topological phase.

We want to understand the general behavior of NNs, so we study multiple instances of CNNs obtained by training them using the same hyperparameters but from different random initializations, following the default uniform initialization in PyTorch \cite{paszke_pytorch_2019}. We train and test our CNNs on data from a system without added disorder, described by \eq~\eqref{eq:SSH_clean_ham}. We train them on data sampled in equal numbers from both topological and trivial phases, far from the phase transition. The loss function we optimize in this task is defined by the binary cross-entropy (BCE) function \cite{good_bce_1952}. We test CNNs on data sampled uniformly over a whole range of $\intra/\inter$, including the close vicinity of the phase transition. The $\intra/\inter$ values used to generate data points in the three data sets (training, validation, and test) were different to prevent possible leakage of information. Such choice of testing points allows us to assess the networks' \textit{generalization}, that is, its performance on previously unseen examples that are sampled from the same underlying distribution of the data. Our understanding is that data that are drawn from the same distribution share the same structure and set of features.

 A much more challenging task for NNs than the in-distribution generalization is to achieve out-of-distribution (OOD) generalization, that is, to perform well on data that come from a different distribution than the training data. We expect ODD generalization when OOD data have a similar structure and set of meaningful features as training data. In our case, these OOD data are data samples constructed from eigenstates of the disordered SSH model described by \eq~\eqref{eq:ssh_disordered_ham}. Many new features are added to the OOD data that arise from the disorder-induced Anderson localization~\cite{Anderson1958absence}. This results in more highly localized eigenstates, which render the edge states less distinguishable with increasing disorder strength. Especially dangerous to the OOD generalization are \textit{spurious correlations}, which can be present in the training data. They refer to a situation where some features of the input data and the label appear to be related to each other, but the relationship is coincidental or confounded by an external variable~\cite{ye2024spurious}.

To evaluate the network OOD generalization, we check its ability to recreate the phase diagram faithfully. A viable error metric is the root mean square error (RMSE). In our case, it is computed as the square root mean of an element-wise square difference between the target ($y$) and the predicted ($\hat{y}$) winding numbers, for $N_\mathrm{s}^\intra$ values of intracell tunneling $\intra/\inter$, $N_\mathrm{s}^W$ values of the disorder amplitude $W/\inter$, and $N_\mathrm{r}$ realizations of each disorder amplitude $W/\inter$.
The number of entries in the $y$ and $\hat{y}$ arrays is therefore, \,$N_\mathrm{el}\,=\,N_\mathrm{s}^W\,\cdot\,N_\mathrm{s}^\intra\,\cdot\,N_\mathrm{r}$, and the RMSE formula reads
\begin{equation}
    \mathrm{RMSE}(y, \hat{y}) = \sqrt{\frac{1}{N_\mathrm{el}}\sum_{i=1}^{N_\mathrm{el}} \qty(y_i - \hat{y}_i)^2}\,.
\end{equation}
Another metric we use is OOD accuracy, defined as the percentage of correct predictions out of $N_\mathrm{el}$ in the winding number prediction task. Note that the network can output only binary predictions, however, when we plot phase diagrams predicted by networks there are non-binary entries coming from averaging over $N_\mathrm{r}$ realizations of each disorder amplitude $W/\inter$.

\subsection{Class activation mapping (CAM)}\label{sec:CAM}

Machine learning methods are usually black-box tools. They accomplish the tasks that we give them at the cost of not being able to justify the outcome they provide. This shortcoming is known as the lack of \textit{interpretability}. Interpretability can be understood as ``the degree to which a human can consistently predict the model's result'' \cite{Kim16ANIPS}. There are numerous interpretability techniques that aim at a better understanding of the network reasoning. We recommend their overview in \reflab~\cite{Molnar19book}. In this work, we apply class activation mapping (CAM), a simple pixel attribution technique. CAM, given an input sample, produces a map highlighting the areas of the input that contribute the most to the prediction of a considered class. 

CAM is an attribution technique tailored for use with CNNs and leverages their design. A CNN produces different representations of the input data during a forward pass through the network and encodes them in different channels. CAM relies on the latent representation of the data present in the output channels of the last convolutional layer in a network, visualized in \fig~\ref{fig:intro}(b) as colored activation maps, 
and performs their weighted sum with weights $\alpha_k$ \cite{jung2021better}.
The resulting map, after rescaling back to the original data size, highlights the areas that were the most influential in predicting a considered class. The crucial part is the choice of the weights $\alpha_k$. The original formulation of the method relies on reducing each channel output of the last convolutional layer to a single number with global average pooling (GAP) layer~\cite{Zhou16CAM}. The weights of fully-connected layer connecting those numbers with the output corresponding to the considered class $C$ are then used as $\alpha_k$ for the weighted average of convolved images (see \app~\ref{app:CAM_is_fragile} for more details). An obvious limitation of this approach is the necessity for the NN architecture to include the GAP layer, which can limit the applicability of CAM to existing trained models.

This technique can be made architecture-agnostic \cite{Smilkov17, Selvaraju17, Chattopadhay18, Desai20, Fu20, Wang20}. The simplest extension is Grad-CAM \cite{Selvaraju17}, which replaces the need for the GAP layer with gradients computed by backpropagation of the class output to the last convolutional layer. 
We applied both techniques in our analysis, but they gave quantitatively the same results, which is why in the discussion we only present the results obtained with CAM.

\subsection{Dimensionality reduction}\label{sec:PCA_UMAP}
The information learned by an NN is stored throughout all its parameters. Analyzing how data is represented in the parameter space could help researchers extract this knowledge. However, even for small NNs, the dimensionality of parameter space is of the order of thousands, and humans have difficulty comprehending data in high dimensions. Thus, reducing data to a small number of dimensions is helpful for visualization purposes and allows us to gain precious insight into NNs’ inner workings~\cite{Erhan09, Simonyan13, Nguyen15, Olah17}. An example of such insight might be to extract which areas in latent space correspond to given learned concepts and, conversely, the relations and distance between them the network has learned~\cite{Carter19}.
To take advantage of this paradigm in our work, we use principal component analysis (PCA) - an established linear dimensionality reduction technique.

PCA~\cite{Pearson1901} relies on computing the principal components (PCs)
of the data that are first stacked to form a multidimensional tensor. The PCs are eigenvectors of the covariance matrix of this tensor, represented schematically as orange arrows in \fig~\ref{fig:intro}(c). The magnitude of their corresponding eigenvalues orders them, so the first PC represents the direction of the highest data variance, the second PC is orthogonal to it and describes the direction with the second highest variance, etc. The original tensor is then projected into a subspace spanned by the selected number of its PCs. The result is a tangible low-dimensional representation of the data that presents the maximal variance of the original high-dimensional space. The low-dimensional representation is needed as a preprocessing step for some ML algorithms or visualization purposes. 
Because of the linearity of this reduction technique, the distance between the points and the density is primarily preserved. It can be treated as representative of the distance in the original high-dimensional space. 

\section{Results and discussion}\label{sec:results_discussion}

\subsection{Convolutional neural networks fail to generalize to data with disorder}\label{sec:CNNs_fail}

Convolutional neural networks succeeded in the training task posed by the disorderless SSH system. The training of all CNN instances converged, and the networks achieved perfect (100\%) accuracy on both the training and validation data sets. The performance was also very good for all the trained instances when tasked with making predictions for in-distribution data. The accuracy score for the in-distribution testing was greater than 95\%. The only regions where some test samples were misclassified were for data in the vicinity of the phase transition, which were excluded from the training set. 

\begin{figure}[t]
    \centering
    \includegraphics[width=\columnwidth]{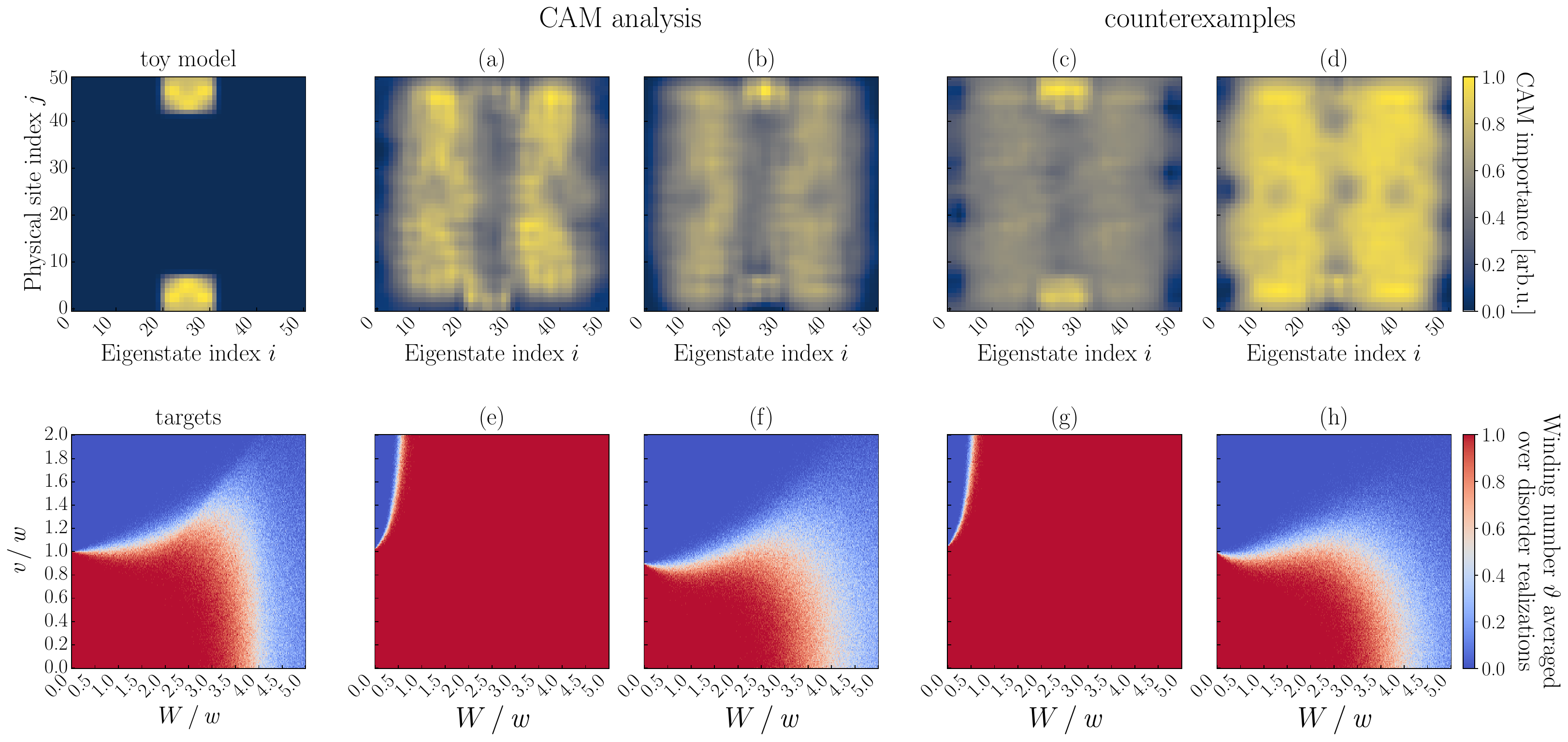}
    \caption{\textbf{Comparison between a well- and poorly-generalizing CNN through the lens of CAM}. First column: the expected CAM result for a CNN that focuses on the edge states, obtained via a toy model in \app~\ref{app:toy-cam} (first row), and the target phase diagram (second row). (a) CAM-based explanation of the prediction of one of many poorly-generalizing CNNs (here, the OOD accuracy of $50.3\%$) for an input sample from the disorderless SSH model at $\intra / \inter=0.27$. CNN places more importance on states around the one- and three-fourths of the spectrum. (b) One of the well-generalizing CNNs (OOD accuracy of $90.1\%$) in the same input sample pays more attention to the localization of the edge states. However, (c) poorly-generalizing CNN (OOD accuracy of $52.3\%$) can also pay attention to edge states, and (d) well-generalizing CNN (OOD accuracy of $91.7\%$) can ignore edge states, rendering the CAM analysis inconclusive. Panels (e)-(h) present the predicted phase diagram of the corresponding CNNs. The phase diagrams are predicted winding numbers $\windnum$ averaged over $N_\mathrm{r}=25$ realizations of disorder, with $N_\mathrm{s}^\intra=200$ slices in $\intra/\inter$, and $N_\mathrm{s}^W=500$ slices in $W/\inter$ directions.}
    \label{fig:cam_and_generalizations}
\end{figure}

Despite promising results within the training distribution, OOD generalization proved to be a challenging task, as we show in \tab~\ref{tab:result_stats}. We quantify the OOD generalization by testing the trained networks on the phase diagram of the SSH model, generated for 25 disorder realizations, and presented in the bottom-left corner of \fig~\ref{fig:cam_and_generalizations}. Upon many initializations, we have found that only as little as $5\%$ of them manage to reproduce the shape of the phase diagram correctly. We used RMSE as a metric of distinction between the well- and poorly-generalizing models, with the threshold set at 0.2. By dividing these models, we can also observe their OOD accuracy. The well-generalizing networks achieve an OOD accuracy of 84\% and more on the data from the whole phase diagram. In contrast, a large majority (95\%) of CNNs have OOD accuracy only marginally higher than a random guess, as seen in the third row of \tab~\ref{tab:result_stats}. 
Interestingly, the well-generalizing networks tend to predict the phase transition at values below $\intra/\inter=1$, closer to $\intra/\inter \approx 0.9 - 0.95$, as visible in \figs~\ref{fig:cam_and_generalizations}(h) and (j). They may take into account the finite-size effects that the infinite-system approximation of the winding number fails to account for.

We can improve this poor statistic by introducing slightly disordered ($W/ \inter \in\qty{0.01, 0.05}$) data to the training data set. We treat this range of small disorders as a 'perturbative' regime in which we can label the slightly disordered data with winding numbers calculated for respective disorderless points of the phase diagram.
As a result, the ratio of successful CNNs increases from $5\%$ to $22\%$. Other statistics remain qualitatively the same in both well- and poorly-generating models. Performance comparison of 100 randomly initialized CNNs trained only on disorderless data and CNNs trained also with slightly disordered data is in \tab~\ref{tab:result_stats}.

The networks exhibit this behavior despite showing stellar performance both in the training regime and during the in-domain testing. Most CNNs fail even though we provide them with all the necessary information to carry this performance to a disordered regime, such as the disorder-robust connection between zero-energy edge states and winding number. Moreover, the physically motivated remedy only slightly improved the number of well-generalizing networks. To understand these phenomena, we must lift the lid of the 'black box' and ask why CNNs do not generalize well.

\definecolor{Silver}{rgb}{0.9,0.9,0.9}
\definecolor{DustyGray}{rgb}{0.8,0.8,0.8}
\begin{table}[b]
\centering
\caption{Statistics (means $\pm$ standard deviations) of 100 randomly initialized networks trained only on data without disorder, $W=0$, and on data including also slightly disordered samples, $W/ \inter \in\qty{0, 0.01, 0.05}$.}
\label{tab:result_stats}
\begin{tblr}{
  width = \linewidth,
  colspec = {Q[185]Q[181]Q[183]Q[192]Q[194]},
  cells = {c},
  cell{1}{2} = {c=2}{0.364\linewidth},
  cell{1}{4} = {c=2}{0.386\linewidth},
  cell{2}{2} = {Silver},
  cell{2}{3} = {DustyGray},
  cell{2}{4} = {Silver},
  cell{2}{5} = {DustyGray},
  cell{3}{2} = {Silver},
  cell{3}{3} = {DustyGray},
  cell{3}{4} = {Silver},
  cell{3}{5} = {DustyGray},
  cell{5}{2} = {Silver},
  cell{5}{3} = {DustyGray},
  cell{5}{4} = {Silver},
  cell{5}{5} = {DustyGray},
  cell{6}{2} = {Silver},
  cell{6}{3} = {DustyGray},
  cell{6}{4} = {Silver},
  cell{6}{5} = {DustyGray},
  cell{7}{2} = {Silver},
  cell{7}{3} = {DustyGray},
  cell{7}{4} = {Silver},
  cell{7}{5} = {DustyGray},
  cell{8}{2} = {Silver},
  cell{8}{3} = {DustyGray},
  cell{8}{4} = {Silver},
  cell{8}{5} = {DustyGray},
  vline{2-3,5} = {1}{black},
  vline{3-5,5} = {1}{black},
  vline{2-6} = {2-3,5-8}{black},
  hline{2,4-5} = {-}{black},
}
 & CNNs trained on \(W=0\) &  & CNNs trained on \(W/ \inter \in\qty{0, 0.01, 0.05}\) & \\
OOD generalization & Well-generalizing & Poorly-generalizing & Well-generalizing & Poorly-generalizing\\
{
Number of CNNs
} & 5\,/\,100 & 95\,/\,100 & 22\,/\,100 & 78\,/\,100\\ \\
Training accuracy & \(\sim 100\,\%\) & \(\sim 100\,\%\) & \(\sim 100\,\%\) & \(\sim 100\,\%\)\\
Test accuracy & \(95.5 \pm 1.5 \, \%\) & \(97 \pm 2 \,\%\) & \(94.7 \pm 1.9 \, \%\) & \(97 \pm 2 \, \%\)\\
OOD accuracy & \(84 \pm 3\,\%\) & \(55 \pm 8\,\%\) & \(86 \pm 3\,\%\) & \(61 \pm 9\,\%\)\\
RMSE & \(0.168 \pm 0.025\) & \(0.46 \pm 0.07\) & \(0.153 \pm 0.031\) & \(0.41 \pm 0.10\)
\end{tblr}
\end{table}

\subsection{Analysis  of the CNNs generalization's failure}\label{sec:why_fail}

Most successfully trained CNNs failed to generalize from the simple training regime to the disordered regime. This is especially surprising because there is a robust indicator of the topological phase, namely the presence of edge states. To understand what other features of the input data are instead leveraged by networks to make their predictions, we use class activation mapping (CAM) \cite{Zhou16CAM}, described in \seclab~\ref{sec:CAM}.

A large majority of trained networks tend to ignore edge states in the middle of the spectrum as well as the extremes of the spectrum, that is, the ground state and the highest excited state (maximally filled state). Instead, they look at the remaining states of the system. We see an example of such a typical network in \fig~\ref{fig:cam_and_generalizations}(a). Networks that ignore edge states in the disorderless SSH model usually fail to generalize to the system with the disorder, as seen in the corresponding predicted phase diagram in \fig~\ref{fig:cam_and_generalizations}(e). In contrast, a small number of networks focus more on the edge states in the middle of the spectrum, corresponding to the half-filled system and constituting the topological invariant of the system. An example of such a network is shown in \fig~\ref{fig:cam_and_generalizations}(b), which pays closer attention to the localization at one of the system edges. This group of networks is more likely to generalize well to the disordered data, see the predicted phase diagram in \fig~\ref{fig:cam_and_generalizations}(f), as it detects something related to the known topological invariant. This analysis shows that we can increase our trust in the OOD generalization of the network by making sure it looks at relevant features in the known regime of the problem.

\begin{figure}[t]
    \centering
    \includegraphics[width=0.9\columnwidth]{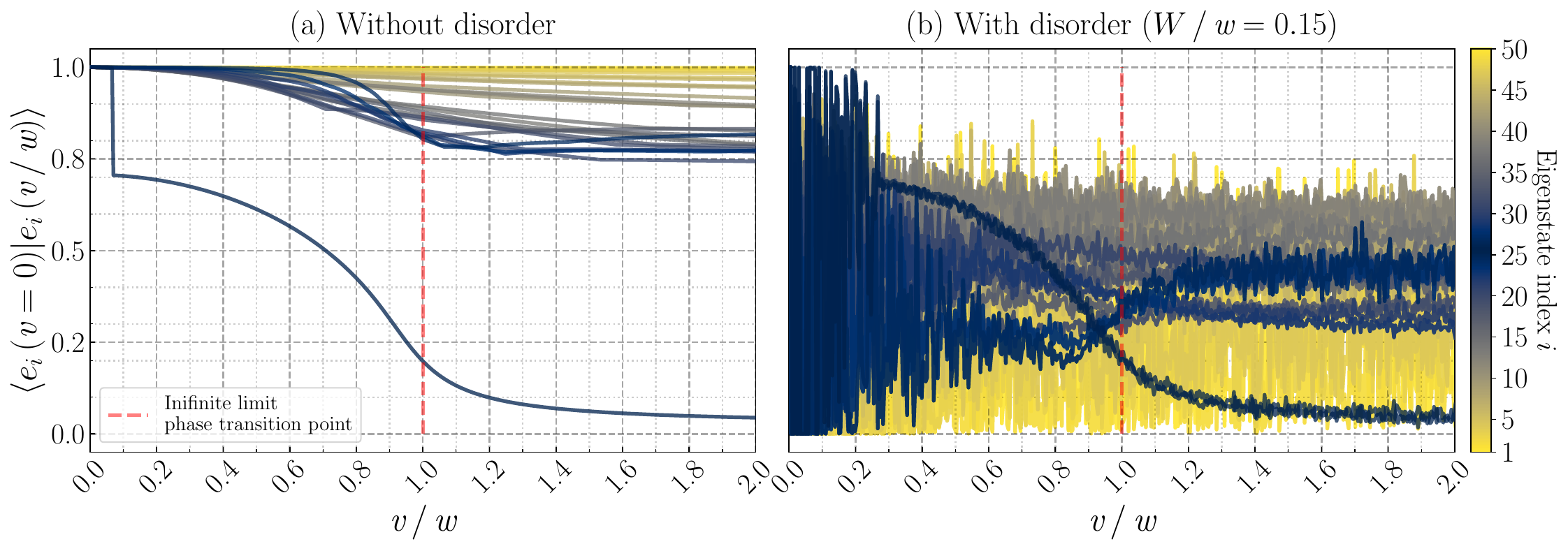}
    \caption{\textbf{Fidelity between the eigenstate $e_i$ for $\intra=0$ and the respective eigenstate $e_i$ for different values $\intra / \inter$}, showcasing an example of correlations in data coming from the SSH model that can be used by an NN to predict a label in a simple regime. (a) Apart from the rapidly changing fidelity expected of the half-filled edge state, we can also observe the changing fidelity of eigenstates closer to the extremes of the spectrum. (b) Already in the weakly disordered case ($W/ \inter =0.15$) fidelity of the states closer to the extremes of the spectrum becomes a poor predictor of the phases. Only the disorder-robust fidelity of edge states remains a valid feature. Note: The drop in fidelity of edge states around $\intra/\inter = 0.08$ in panel (a) and rapid oscillations for small $\intra/\inter$ in panel (b) come from the hybrydization of the edge states due to finite-size effects \cite{asboth_short_2016}.}
    \label{fig:correlations}
\end{figure}

The analysis so far showed that networks tend to ignore edge states, but why is this the case? We hypothesize that features must exist in the bulk that allow a network to solve the learning task in the simple regime, i.e., that correlate well with the label in the simple regime. Prompted by this analysis, we compare each eigenstate of the SSH model for intracell tunneling $\intra = 0$ to its respective eigenstate for different intracell tunneling values. We plot the described fidelity of the pair of eigenstates in \fig~\ref{fig:correlations}(a). As expected, we see that the fidelity of the edge states at $\intra=0$ and their counterparts for different $\intra/\inter$ changes rapidly throughout the phase diagram. However, we observe that the fidelity of the bulk states also changes across $\intra/\inter$ and, as such, can be used by a CNN to predict a label in the system without disorder. It is only an example of a possible feature, and there can be any number of more complex ones that the networks pick up from the data. Importantly, they no longer correlate with the label in the disordered system, as seen in \fig~\ref{fig:correlations}(b). For this reason, we consider them to be \textit{spurious correlations} in the task of distinguishing between the topological and topologically trivial phases.

Apparently, in this task, CNNs tend to learn some combination of features related to numerous non-edge states, even if they are weakly correlated with the label when taken separately. This observation echoes the results of the ML community on the inherent trade-off of classifiers between the accuracy and robustness. Tsipras \textit{et al.} \cite{tsipras_robustness_2020} showed rigorously on a simple example that classifiers learn a combination of weakly correlated features to achieve perfect accuracy, instead of relying on a single strongly correlated feature that is also present in the data but does not allow for perfect accuracy. When disorder enters the data, it can easily distort the combination of weakly correlated features, harming robustness of a classifier. A solution to achieving robust neural networks is adversarial training \cite{madry2018towards}, which is reminiscent of the original solution from \reflab~\cite{huembeli_identifying_2018}, which used a domain adversarial neural network. In our data distribution, though, the single feature (edge states) should be perfectly correlated with the label (up to finite-size effects), so the task of maximizing accuracy should not lead to ignoring this feature. We do not know what causes CNNs to favor the combination of weak correlators over the single predictive feature, but we have two hypotheses. First, the preference towards multiple features may be due to regularization that prevents NNs from relying on single features. Second, it may come from the iterative nature of the training, which causes multiple features with weak correlation to have an increasingly strong training signal as they boost each other.

\subsection{Unsupervised analysis of OOD generalizations of NNs}
\label{sec:OOD_howto}

In the previous section, we have described how CAM guided us into understanding that the majority of CNNs learn some spurious combination of weakly correlated features (related to non-edge states) instead of a single strong feature (related to edge states). An appropriate combination of weakly correlated features allows for perfect accuracy in the simple regime but ceases to correlate with the label in the disordered regime. The same CAM analysis can be used to assess the OOD generalization of CNNs without access to the labels. For example, if a CNN focuses on the edge state, this increases the chance of a better OOD generalization of the model.

Although, overall, CAM is useful and can be used as the first step of the generalization study, we have found that the CAM analysis is unfortunately inconclusive and can only indicate CNN OOD generalization tendencies. In particular, not every CNN that pays attention to the edge states exhibits good OOD generalization -- evidence for this is in panels (c) and (g) of \fig~\ref{fig:cam_and_generalizations}. We even find an opposite example where the CNN that does not put special focus on edge states achieves great OOD accuracy, as seen in panels (d) and (h) of \fig~\ref{fig:cam_and_generalizations}. We share the CAM results for all 100 trained CNNs for all test data online~\cite{noauthor_github_2023} 
for the reader interested in various possible behaviors of the discussed CNNs. The reliability of CAM as the method to assess the generalization could be improved if used on the test data from the disordered SSH model. However, this would bring this explanation method into a noisy regime where CAM is known to be fragile and unreliable. We advise the reader against using this explanation method in the presence of noise and elaborate on this topic in \app~\ref{app:CAM_is_fragile}.

As a second technique to validate the OOD generalization of a network, we propose to study the structure of data representation learned by trained CNNs. When an NN processes an input sample, it generates different activations at every network layer. We want to understand the relation between the activations generated by data without the disorder and those generated by the disordered data. We follow here the logic that data the network views as similar should generate similar activations or, in other words, should have similar learned representations. If an NN sees any meaningful relation between the training data without the disorder and the OOD test data with the disorder, their representations should be somewhat similar. Such a result increases our trust in the OOD prediction made by a network.

The space of CNN activations is high-dimensional. In the case of a fully connected layer, the dimension depends on the number of neurons. In the case of a convolutional layer, the dimension is equal to the size of the convolved data point. To study the relation between points in such a high-dimensional space, we apply dimensionality reduction techniques that can bring the space dimension down to, e.g., two while preserving some structure of the original data. We use here principal component analysis (PCA), explained in \seclab~\ref{sec:PCA_UMAP}. To verify the results, we also used uniform manifold approximation and projection (UMAP), which is a non-linear dimensionality reduction technique. We discuss obtained results in \app~\ref{app:evolution_across_layers}.

\begin{figure}[t]
    \centering
    \includegraphics[width=\columnwidth]{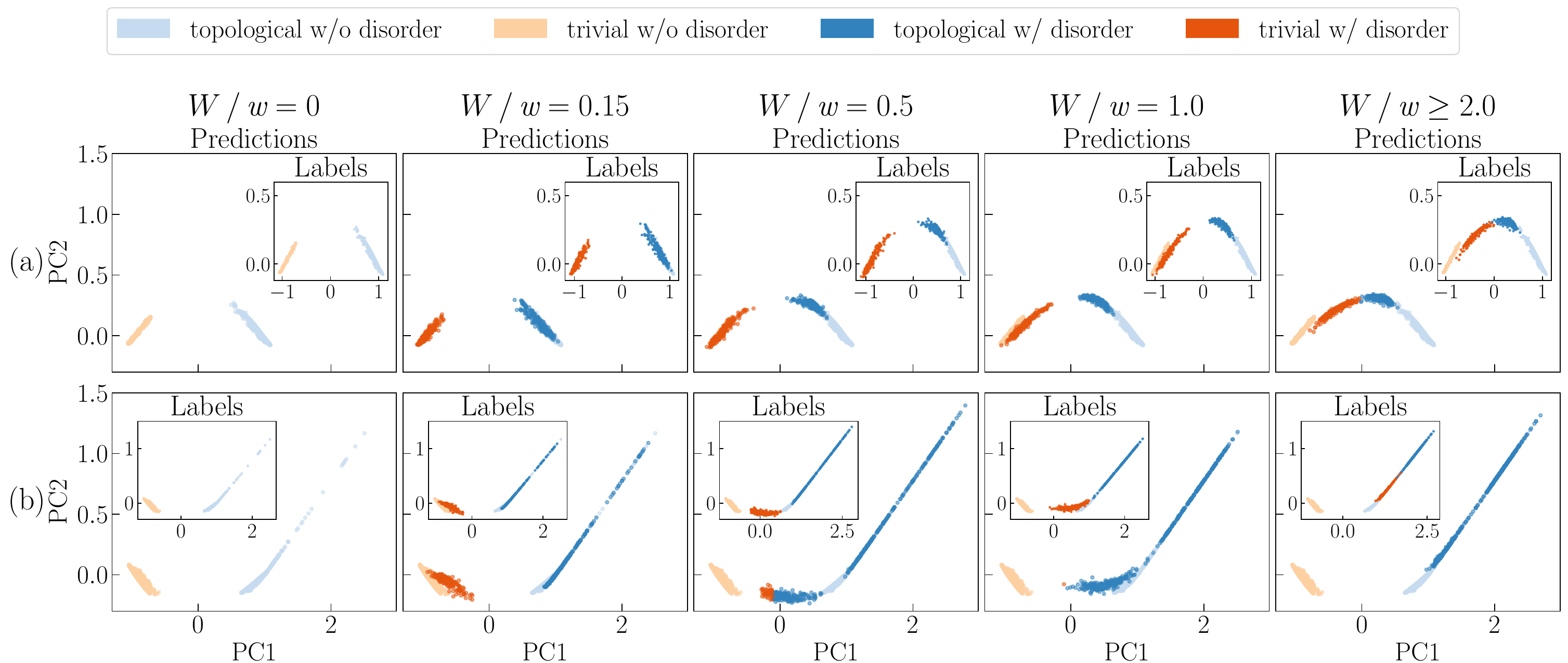}
    \caption{\textbf{Two-dimensional representation of the data learned by the penultimate layer of a poorly- and well-generalizing CNN} with inconclusive CAM results from \fig~\ref{fig:cam_and_generalizations}, obtained with PCA. 
    (a) The well-generalizing CNN, which according to CAM ignores the edge states, learned the representation in which the disorderless data (first column) are clustered in two well-separated clusters corresponding to two phases. Data with increasing disorder falls firstly in the same two clusters formed by disorderless data, and then slowly separates from them but keeps a weak connection.
    (b) The representation of test data learned by the poorly-generalizing CNN, which according to CAM pays close attention to the edge states, already for disorderless data (first column) lacks well-defined two clusters. The data with disorder, forms more and more separated third cluster, only weakly connected to one of the clusters of disorderless data.
    Insets represent labels obtained with the infinite-system approximation of the winding number.
    }
    \label{fig:clustering_final_layer}
\end{figure}

We study the representation of the data learned by the penultimate layer of CNN, that is, the neuron activations after applying the GAP and before entering the last fully connected layer, with the softmax function, which plays the role of a classifier.
To this end, in \fig~\ref{fig:clustering_final_layer} we present the clustering obtained with PCA for data with different disorder strengths in the two-dimensional versions of the high-dimensional representation space of the penultimate CNN layer. The dimensionality reduction is applied to the activations generated by all test data simultaneously, but we plot data subsets for different $W/\inter$'s in separate subplots to make the analysis clearer.
We exclude test data without disorder coming from the vicinity of the transition ($\intra/\inter \in [0.9; 1.1]$) to further simplify the analysis.

Let us first focus on the data representation learned by the penultimate layer of a well-generalizing CNN in  \fig~\ref{fig:clustering_final_layer}(a). The first thing to notice is the representation of the disorderless data (first column) that relies on two well-separated clusters corresponding to two phases (light orange is trivial, light blue is topological). In the next columns, we plot both the disorderless data and the data with increasing disorder strength (indicated with darker orange and blue). For small and intermediate disorder ($W/ \inter = 0.15$ and $0.5$), the disordered data are represented very similarly to the data without disorder, in other words, their representation overlaps. With increasing disorder ($W/ \inter = 2$ and more), the data start to disconnect from the original clusters. Interestingly, the topological data with disorder separate faster from the disorderless data than the data from the topologically trivial phase. Finally, only for large disorder the data form a separate third cluster. This means that the network has a disjointed representation only of strongly disordered data. 

In contrast, the data representation learned by poorly-generalizing CNNs already at the level of disorderless data do not form two well-defined clusters corresponding to two phases. Especially the topological data without disorder form quite a disconnected cluster, as seen with PCA in \fig~\ref{fig:clustering_final_layer}(b). The data with disorder of $W/ \inter =0.5$ (third column) is already separate from the original clusters. For strong disorders, data form three or four separate clusters in the representation space of the penultimate layer. This suggests that the network does not see continuity in the data as a function of the disorder strength.

Notably, the two CNNs whose learned data representations we studied above are the same CNNs whose CAM analysis rendered inconclusive results. That is, the well-generalizing CNN is the one from \fig~\ref{fig:cam_and_generalizations}(d),(h) that focuses less on edge states. The poorly-generalizing CNN is the one from \fig~\ref{fig:cam_and_generalizations}(c),(g) that focuses prominently on edge states. This shows that data representation analysis together with CAM can make stronger statements about the OOD generalization of trained networks.

Finally, we make an analogous analysis for data representation across layers of the well- and poorly-generalizing CNNs in \app~\ref{app:evolution_across_layers}, this time with UMAP. It renders the same conclusions as PCA. However, we make an additional interesting observation that UMAP performed on the input data with and without disorder does better job in forming clusters corresponding to topological and trivial phases than a majority of CNNs trained in a supervised way on data without disorder.

\section{Conclusion and outlook}\label{sec:conclusion}

In this work, we have presented how to increase the trust in predictions of a neural network (NN) when the predictions are made on data from a different distribution than the NN's training data distribution and without access to ground-truth labels. Such an out-of-distribution (OOD) generalization is a desired property of a robust and reliable network and is difficult to achieve or validate. We have shown how one can study the network to understand the way it processes the data with two different tools. The first is an interpretability technique that highlights parts of the data that were important to the network classification decision, called class activation mapping (CAM). The second is an analysis of the data representation learned by the network, facilitated by the dimensionality reduction techniques, such as principal component analysis (PCA), applied to the representation space of the network. The data we have tackled are eigenstates of the Su-Schrieffer-Heeger (SSH) Hamiltonian, coming from two regimes: with and without disorder. We trained hundreds of convolutional networks (CNNs) on the data without disorder and checked their in-distribution and out-of-distribution generalization to data without and with disorder, respectively.
We have made the following observations.
\begin{itemize}
    \item An overwhelming majority ($95\%$) of successfully trained CNNs failed to generalize to data with disorder. Expanding the training set with data with slight disorder ($W/ \inter \leq 0.05$) improved these statistics only to an extent ($78\%$).
    \item The reason CNNs tend to fail is that they focus on non-generalizable features of data that are nevertheless useful for the training task, as indicated by the CAM analysis of CNNs' predictions. For example, eigenvectors other than the edge states carry correlations that can be leveraged to classify phases in the disorderless data regime, but they disappear when disorder is added.
    \item CAM analysis can be used to assert our trust in network prediction. If in the disorderless regime a network pays attention to edge states, which are known to be useful features, such a network tends to exhibit a better OOD generalization. Surprisingly, such networks can still fail to generalize to disordered data, making the CAM analysis inconclusive.
    \item Moreover, CAM itself is also a fragile interpretation technique that performs poorly in the presence of noise, which prevented its use to understand the network predictions in the disordered regime.
    \item Dimensionality reduction techniques such as PCA can be used to visualize the data representation learned by networks. If a CNN has a disconnected (very different) representation of disordered data with no connection to the disorderless training data, the trust in its predictions on the disordered data should be limited. A good predictor of the OOD generalization quality is instead when a CNN represents the data with slight disorder similarly to its disorderless training data and then the representation gets smoothly disconnected with increasing disorder strength.
\end{itemize}

We conclude that, together, CAM and data representation analysis serve as useful tools to gain additional insight and assess trust in NN predictions. Given their low cost, we believe that their routine use can only benefit members of the scientific community who have already added deep learning to their computational toolbox. Such an analysis of NNs is especially needed in light of our surprising observations that CNNs fail in a relatively simple task that is generalizing to slight disorder when trained on the disorderless SSH model, whereas physicists know that there exists a robust feature to solve this task, such as the presence of edge states. This observation also highlights that scientific data sets create new challenges for deep learning, which can unravel unexpected behaviors and failure modes of NNs. At the same time, the community needs to develop more robust tools to assess NN performance in regimes without known ground truths, bearing in mind that some of them fail in the presence of random or adversarial noise.

\subsection*{Data availability}
The source code and data that support the findings of this study is openly available at \url{https://github.com/kcybinski/Interpreting_NNs_for_topological_phases_of_matter} \cite{noauthor_github_2023}. 

\begin{acknowledgments}
K.C. acknowledges the financial support from the Polish Ministry of Science and Higher Education within the ``Excellence initiative – research university'' program.
M.T. acknowledge the National Science Centre Poland (grant no.~2020/38/E/ST2/00564) for the financial support and the Poland’s high-performance computing infrastructure PLGrid (HPC Centers: ACK Cyfronet AGH) for providing computer facilities and support (computational grant no.~PLG/2023/016878). 
ICFO group acknowledges support from European Research Council AdG NOQIA; MCIN/AEI (PGC2018-0910.13039 /501100011033,  CEX2019-000910-S/10.13039/501100011033, Plan National FIDEUA PID2019-106901GB-I00, Plan National STAMEENA PID2022-139099NB, I00, project funded by MCIN/AEI/10.13039/501100011033 and by the “European Union NextGenerationEU/PRTR" (PRTR-C17.I1), FPI); QUANTERA MAQS PCI2019-111828-2);  QUANTERA DYNAMITE PCI2022-132919, QuantERA II Programme co-funded by European Union’s Horizon 2020 program under Grant Agreement No 101017733); Ministry for Digital Transformation and of Civil Service of the Spanish Government through the QUANTUM ENIA project call - Quantum Spain project, and by the European Union through the Recovery, Transformation and Resilience Plan - NextGenerationEU within the framework of the Digital Spain 2026 Agenda; Fundació Cellex; Fundació Mir-Puig; Generalitat de Catalunya (European Social Fund FEDER and CERCA program, AGAUR Grant No. 2021 SGR 01452, QuantumCAT \ U16-011424, co-funded by ERDF Operational Program of Catalonia 2014-2020); Barcelona Supercomputing Center MareNostrum (FI-2023-3-0024); Funded by the European Union. Views and opinions expressed are however those of the author(s) only and do not necessarily reflect those of the European Union, European Commission, European Climate, Infrastructure and Environment Executive Agency (CINEA), or any other granting authority.  Neither the European Union nor any granting authority can be held responsible for them (HORIZON-CL4-2022-QUANTUM-02-SGA  PASQuanS2.1, 101113690, EU Horizon 2020 FET-OPEN OPTOlogic, Grant No 899794),  EU Horizon Europe Program (This project has received funding from the European Union’s Horizon Europe research and innovation program under grant agreement No 101080086 NeQSTGrant Agreement 101080086 — NeQST); ICFO Internal “QuantumGaudi” project; European Union’s Horizon 2020 program under the Marie Sklodowska-Curie grant agreement No 847648; “La Caixa” Junior Leaders fellowships, La Caixa” Foundation (ID 100010434): CF/BQ/PR23/11980043.
Al.D. acknowledges the financial support from a fellowship granted by la Caixa Foundation (ID 100010434, fellowship code LCF/BQ/PR20/11770012). 
An.D. acknowledges the financial support from the Foundation for Polish Science. The Flatiron Institute is a division of the Simons Foundation.
\end{acknowledgments}

\newpage
\clearpage
\appendix

\section{Alternate Su-Schrieffer-Heeger model formulations}
\label{app:formulations}
The basic formulations of the SSH model \cite{su_solitons_1979} we consider is discussed by Asboth et al. \cite{asboth_short_2016}:
\begin{equation}
    \hat{H} = \intra \sum_{m=1}^N \ketbra{m}\otimes \hat{\sigma}_x + \inter \sum_{m=1}^{N-1} \qty(\ketbra{m+1}{m} \otimes \frac{\hat{\sigma}_x + \i \hat{\sigma}_y}{2} +\text{h.c.})\,.
\end{equation}
In $\mbc{i}^{(\dagger)} = \qty(\mbc{A, i}^{(\dagger)}, \mbc{B, i}^{(\dagger)})$ operators formulation it is,
\begin{equation}
    \hat{H} = \intra \sum_{n=1}^N \mbcd{n} \hat{\sigma}_x \mbc{n} + \inter \sum_{n=1}^{N-1} \qty(\mbcd{n}\frac{\hat{\sigma}_x + \i \hat{\sigma}_y}{2}\mbc{n+1} +\text{h.c.})\,.
\end{equation}

Other works we used for comparison of our results formulate it differently:
\begin{enumerate}
    \item Mondragon et al. \cite{mondragon-shem_topological_2014}:
    \begin{equation}
    \hat{H}_{SSH} = \sum_n \qty{ \sshintra_n \mbcd{n} \hat{\sigma}_y \mbc{n} + \sshinter_n  \qty(\mbcd{n} \frac{\hat{\sigma}_x + \i \hat{\sigma}_y}{2} \mbc{n+1} +\text{h.c.})},
    \end{equation}
    \item Meier et al. \cite{meier_observation_2018}:
    \begin{equation}
    \hat{H}_{SSH} = \sum_n \qty{ \sshintra_n \mbcd{n} \hat{\sigma}_x \mbc{n} + \sshinter_n  \qty(\mbcd{n+1} \frac{\hat{\sigma}_x - \i \hat{\sigma}_y}{2} \mbc{n} +\text{h.c.})},
    \end{equation}
    \item Le et al. \cite{le_topological_2020}:
    \begin{equation}
    \hat{H}_{SSH} = \sum_{j, \sigma={A, B}} \qty{-J\qty[1 + (-1)^{j} \Delta t] \, \mbcd{j+1, \sigma} \mbc{j, \sigma} + \text{h.c.}}.
    \end{equation}
\end{enumerate}

Equations drawn from \reflabs~\cite{mondragon-shem_topological_2014} and \cite{meier_observation_2018}, allow for the presence of disorder in the system through $t_n$ and $m_n$ tunneling amplitudes, corresponding to the same \stress{inter-} or \stress{intra-} cell tunneling as $\inter$ and $\intra$. The correspondence is:
\begin{itemize}
    \item intercell tunneling $\inter \to\, \sshinter_n$,
    \item intracell tunneling $\intra \to\, \sshintra_n$.
\end{itemize}

\begin{figure}[b]
    \centering
    \includegraphics[width=\columnwidth]{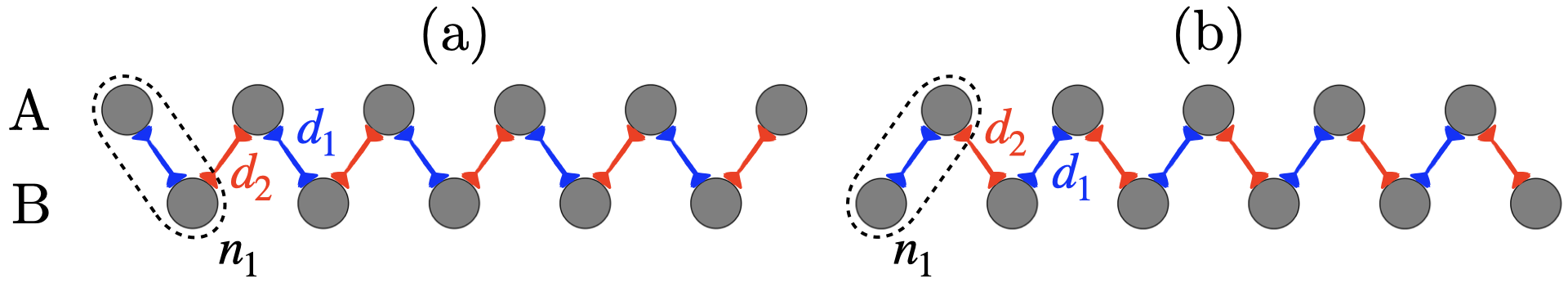}
    \caption{Difference between SSH formulations from \reflabs~\cite{mondragon-shem_topological_2014} and \cite{meier_observation_2018} is due to different sublattice order, as seen in panels (a) and (b), respectively. Both define distance within unit cell $d_1 = 0$, and distance between unit cells as $d_2 = 1$. Clarification of this choice is needed for the further transition to momentum space, as the ordering of A and B sites in each unit cell is swapped.}
    \label{fig:SSH_diff_notation}
\end{figure}

The sign discrepancy between models from \reflab~\cite{mondragon-shem_topological_2014} and \reflab~\cite{meier_observation_2018} in the intercell part is due to the different site numeration. The former uses $BABABA\dots$ convention, and the latter uses $ABABAB\dots$ convention, which is the one we use. The difference is seen in \fig~\ref{fig:SSH_diff_notation}. Another difference we note is that in \reflab~\cite{mondragon-shem_topological_2014}, the intracell tunneling amplitudes are purely imaginary $\sshintra_n$, while in \reflab~\cite{meier_observation_2018}, they are purely real $\sshintra_n$. Both yield the same results as long as one convention is kept. In this work, we opt for real tunneling amplitudes.

\section{Calculation of winding number}
\label{app:windnum}
\textbf{Calculation demonstration.}
The routine we employ for the calculation of the winding number \cite{fraxanet_topological_2021} involves several steps. This section is to serve as a step-by-step demonstration. It is accompanied by a mirror Jupyter Notebook in our GitHub repository~\cite{noauthor_github_2023}. 

\textbf{Demonstration parameters.} In order to keep the calculations tractable, we choose system size to be 6 sites (3 unit cells). We set the boundary conditions to periodic. We generate the occupational basis, and sort it lexicographically.  

\textbf{Winding number formula}
We want to use here the equation for the winding number in the form \cite{fraxanet_topological_2021}:  
\begin{equation} 
    \label{eq:windnum_app}
    \windnum = \frac{1}{2 \pi i} \oint_{BZ} \mathrm{Tr}(h^{-1} \partial_k h) 
\end{equation}
It is defined as a $1^\text{st}$ Brillouin zone integral of a transformed momentum space Hamiltonian. In the following sections we will derive it, starting from Hamiltonian definition presented in \eq~\eqref{eq:SSH_clean_ham}.

\textbf{Occupational basis Hamiltonian.} First we define the Hamiltonian in the occupational basis according to \eq~\eqref{eq:SSH_clean_ham}. In matrix form it is, 
\begin{equation}
    H_{\mathrm{occ}} = \mqty[
        0 & \intra & 0 & 0 & 0 & \inter\\
        \intra & 0 & \inter & 0 & 0 & 0\\
        0 & \inter & 0 & \intra & 0 & 0\\
        0 & 0 & \intra & 0 & \inter & 0\\
        0 & 0 & 0 & \inter & 0 & \intra\\
        \inter & 0 & 0 & 0 & \intra & 0
    ].
\end{equation}

\textbf{Momentum basis Hamiltonian.} The winding number is defined as a $1^\text{st}$ Brillouin zone integral, so we must transfer the Hamiltonian to momentum space. The condition to be met here is that a fermion only accumulates phase when crossing between the neighboring Brillouin zones. Given our choice of boundary conditions, this is achieved by addition of $e^{\i k}$ multiplication to the corner terms. These are the terms corresponding to transitioning the periodic boundary. The resulting Hamiltonian is 
\begin{equation}
    H_{\mathrm{occ}}^\mathrm{k} = \mqty[
        0 & \intra & 0 & 0 & 0 & \inter e^{\i k}\\
        \intra & 0 & \inter & 0 & 0 & 0\\
        0 & \inter & 0 & \intra & 0 & 0\\
        0 & 0 & \intra & 0 & \inter & 0\\
        0 & 0 & 0 & \inter & 0 & \intra\\
        \inter e^{\i k} & 0 & 0 & 0 & \intra & 0
    ].
\end{equation}

\textbf{Chiral symmetry operator.} The next step is to define the chiral symmetry operator $\Gamma$. In this system it has the form
\begin{equation}
    \Gamma = \left[
        \begin{matrix}
            1 & 0 & 0 & 0 & 0 & 0\\
            0 & -1 & 0 & 0 & 0 & 0\\
            0 & 0 & 1 & 0 & 0 & 0\\
            0 & 0 & 0 & -1 & 0 & 0\\
            0 & 0 & 0 & 0 & 1 & 0\\
            0 & 0 & 0 & 0 & 0 & -1
        \end{matrix}
    \right].
\end{equation}

\textbf{Hamiltonian in chiral symmetry basis.}
We need to rewrite our Hamiltonian in the eigenbasis of $\Gamma$. To this end, we must ensure the eigenvectors of~$\Gamma$ are also sorted lexicographically. The resulting change-of-basis matrices~$\chi, \chi^\dagger$ have the form
\begin{equation}
    \chi = \left[
        \begin{matrix}
            0 & 0 & 0 & 0 & 0 & 1\\
            0 & 0 & 0 & 1 & 0 & 0\\
            0 & 1 & 0 & 0 & 0 & 0\\
            0 & 0 & 0 & 0 & 1 & 0\\
            0 & 0 & 1 & 0 & 0 & 0\\
            1 & 0 & 0 & 0 & 0 & 0
        \end{matrix}\right] \qc 
    \chi^\dagger = \left[
        \begin{matrix}
            0 & 0 & 0 & 0 & 0 & 1\\
            0 & 0 & 1 & 0 & 0 & 0\\
            0 & 0 & 0 & 0 & 1 & 0\\
            0 & 1 & 0 & 0 & 0 & 0\\
            0 & 0 & 0 & 1 & 0 & 0\\
            1 & 0 & 0 & 0 & 0 & 0
        \end{matrix}\right].
\end{equation}
Once we apply them to the Hamiltonian $H$ it becomes block off-diagonal, 
\begin{equation}
    H = \chi H_{\mathrm{occ}}^\text{k} \chi^\dagger = \left[
        \begin{matrix}
            0 & 0 & 0 & v & 0 & we^{-\i k}\\
            0 & 0 & 0 & w & v & 0\\
            0 & 0 & 0 & 0 & w & v\\
            v & w & 0 & 0 & 0 & 0\\
            0 & v & w & 0 & 0 & 0\\
            we^{\i k} & 0 & v & 0 & 0 & 0
        \end{matrix}
        \right]
     = \mqty[0 & h^\dagger \\ h & 0].
\end{equation}

\begin{figure}[t]
    \centering
    \includegraphics[width=\textwidth]{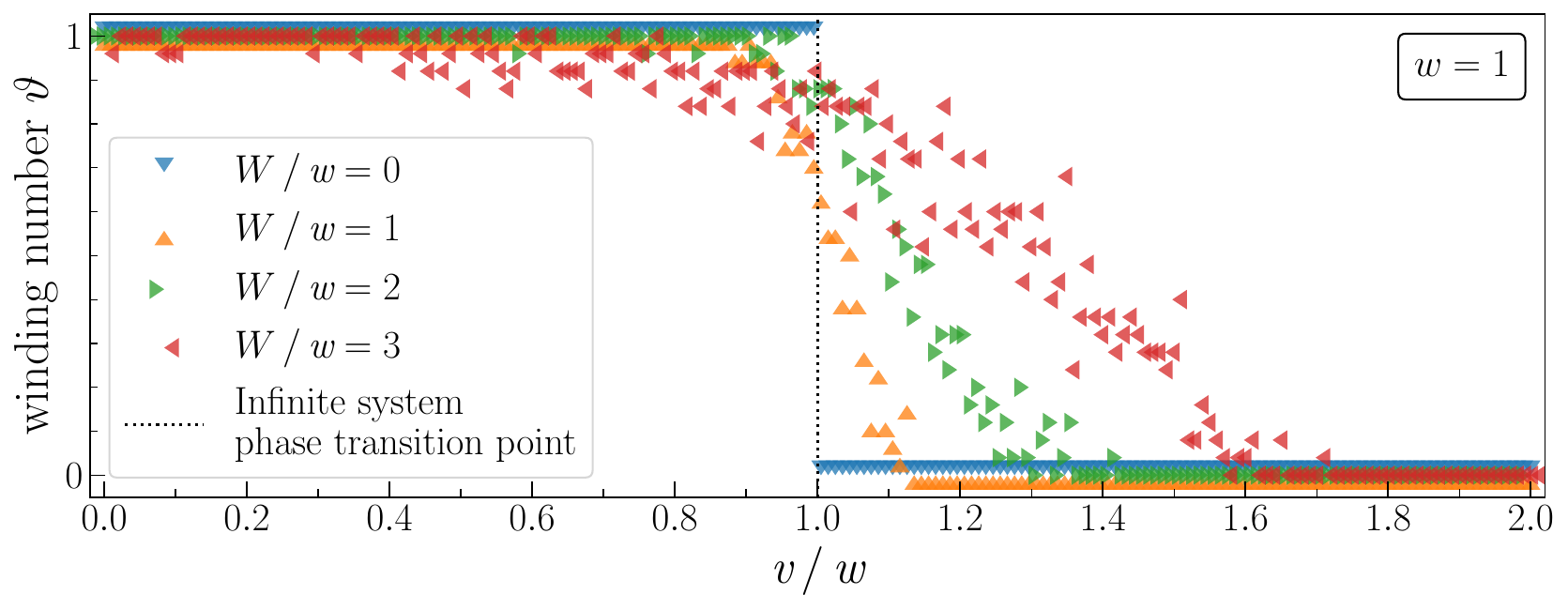}
    \caption{\textbf{Phase transition vs disorder strength.} Comparison of winding number computed for different disorder amplitudes $W/\inter$, averaged over $N_\mathrm{r}=25$ realizations of disorder. With increased disorder, the phase transition point shifts to $\intra/\inter \geq 1$, as discussed in \seclab~\ref{sec:Hamiltonian}. Use of infinite system calculation is visible by the position of phase transition point, computed to happen at $\intra/\inter =1$ in disorderless system ($W=0$). In a finite system, it would happen for a smaller $\intra/\inter$ ratio \cite{asboth_short_2016}.}
    \label{fig:comparison_SSH}
\end{figure}

\textbf{Final calculation.}
Now the elements we need for the calculations according to \eq~\eqref{eq:windnum_app} have the form:
    \[
        h = 
        \left[
            \begin{matrix}
                v & w & 0\\
                0 & v & w\\
                w e^{\i k} & 0 & v
            \end{matrix}
        \right] \qc
        h^{-1} = \frac{1}{v^{3} + w^{3} e^{\i k}}
        \left[
            \begin{matrix}
                v^{2} & - v w & w^{2}\\
                w^{2} e^{\i k} & v^{2} & - v w\\
                - v w e^{\i k} & w^{2} e^{\i k} & v^{2}
            \end{matrix}
        \right] \qc 
        \partial_k h = \left[\begin{matrix}0 & 0 & 0\\0 & 0 & 0\\\i w e^{\i k} & 0 & 0\end{matrix}\right],
    \]
    \[
    h^{-1} \pd{k} h = \frac{ie^{\i k}}{v^{3} + w^{3} e^{\i k}}
    \left[
    \begin{matrix}
        w^{2} & 0 & 0\\
        - v w & 0 & 0\\
        v^{2} & 0 & 0
    \end{matrix}
    \right]
    \qc
        \tr \qty[h^{-1} \pd{k} h] = \frac{iw^2e^{\i k}}{v^{3} + w^{3} e^{\i k}}.
    \]

    We have arrived at the final formula for integration to obtain the winding number. Now, it needs to be integrated over the $1^\text{st}$ Brillouin Zone. Due to software stability reasons, we integrated it numerically for $k \in \qty[0, 2\pi - \epsilon]$. 
    
    \textbf{Universality.} This calculation of the winding number also applies to the disordered SSH Hamiltonian, defined by \eq~\eqref{eq:ssh_disordered_ham}. The phase transition point then drifts to values of $\intra/\inter \geq 1$ with an increase of disorder amplitude $W/\inter$. This is presented in \fig~\ref{fig:comparison_SSH}.

\section{Architectures and hyperparameters}
\label{app:architectures}
\textbf{Data point.} Single data point supplied to the network is a matrix of squared coefficients of eigenstates 
of Hamiltonians defined by \eqs~\eqref{eq:SSH_clean_ham} (represented in a lexicographically sorted occupational basis), \eqref{eq:ssh_disordered_ham} and its corresponding label. Each column is a single normalized eigenstate. The label is the winding number. A step-by-step demonstration of the calculation of the winding number is presented in \app~\ref{app:windnum}.

\textbf{Basic training parameters.} All throughout this work we keep the intercell tunneling amplitude $\inter = 1$. The two phases of the SSH system are probed by varying $\intra/\inter\in\qty(0,2)$. In all data sets (training, validation, test) both phases are equally numerous -- 50\% of all data points represent each phase. To ensure linear independence of data in all three sets, the intervals of $\intra/\inter$ were chosen with a slightly offset initial point. That is $0.001, 0.002, 0.003$ for training, validation, and test set, respectively. This ensures they all represent distinct points in the phase diagram, and that in disorderless setting they belong to the same distribution.

\textbf{Data set composition.}
A part of what we would like to achieve is for the network to learn the correct phase transition point by itself. Therefore, the training and validation data sets do not contain data from the direct vicinity of the phase transition. In these two sets, we vary the parameter $\intra$ in the range $\intra / \inter \in \qty(0, 0.8) \cup \qty(1.2, 2)$. The composition of disorderless data sets is presented in the left column of Table~\ref{app:tab:datasets}.

\textbf{Training extension -- Disordered data.} The extension that proved fruitful was an addition of disordered data to the training. In this setting, we included data from low-disorder regimes ($W/\inter \in{0.01, 0.05})$ in training and validation sets. For each disorder amplitude, the data added to the training set remained in~$5:1$ ratio to the disorderless data. For the validation set, the ratio was set at~$4:1$. The composition of data sets used in both regimes of training are presented in Table~\ref{app:tab:datasets}.

\textbf{Generalization assessment.} To assess the networks' generalization, we tasked it with recreating the target~$\intra$~vs~$W$ phase diagram. To this end, we generated~$N_\mathrm{r} = 25$ disorder realizations of the input data and corresponding labels. We covered the range~$\intra / \inter \in \qty[0,2]$, and~$W / \inter \in\qty[0,5]$ for each realization. The step size in disorder amplitude strength is ~$\Delta W / \inter = 0.1$, while the step size in intracell tunneling amplitude is $\Delta\intra / \inter = 0.01$. After collecting predictions for all $N_\mathrm{r}$ realizations, the values in the generated phase diagram are then averaged over $N_\mathrm{r}$ predicted labels for each pair of $\qty(\intra / \inter, W / \inter)$.

\begin{table}[t]
\caption{Composition of data sets used in networks' training. The base setting involved only using data coming from the SSH model in \eq~\eqref{eq:SSH_clean_ham}. In an extension to this approach, data from disordered SSH model was added, see \eq~\eqref{eq:ssh_disordered_ham}.}
\label{app:tab:datasets}
\begin{tabular}{c||c||ccc}
Approach      & Disorderless & \multicolumn{3}{c}{Disordered data}                             \\ \hline \hline
Disorder & \(W=0\) & \multicolumn{1}{c|}{\(W=0\)} & \multicolumn{1}{l|}{\(W / \inter =0.01\)} & \(W / \inter =0.05\) \\ \hline
Training points   & 5000  & \multicolumn{1}{c|}{5000} & \multicolumn{1}{c|}{1000} & 1000 \\ \hline
Validation points & 1000  & \multicolumn{1}{c|}{1000} & \multicolumn{1}{c|}{250}  & 250 
\end{tabular}
\end{table}

\begin{figure}[t]
\begin{center}
\includegraphics[width=0.99\columnwidth]{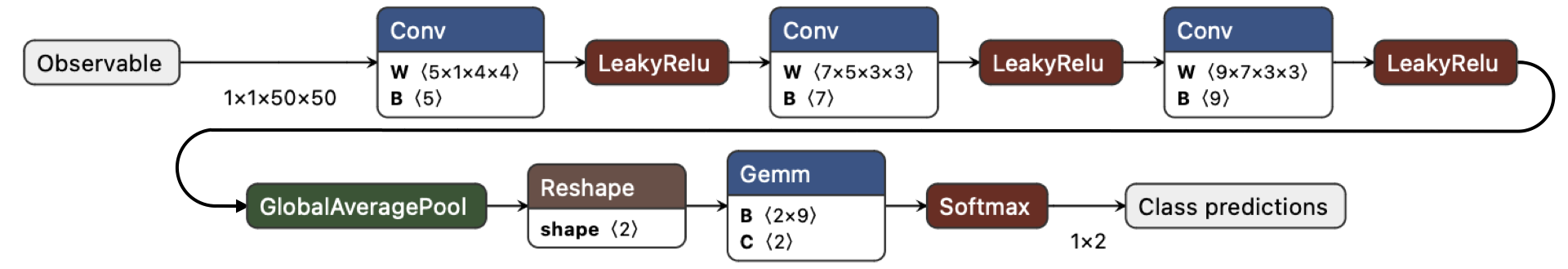}
\end{center}
\caption{Architecture of the CNN used in this work.}
\label{fig:app:architecture}
\end{figure}

\textbf{Architecture.}
The architecture we use is a shallow convolutional neural network (CNN), as presented in \fig~\ref{fig:app:architecture}. It consists of three convolutional layers with symmetric kernels. The initial convolution filter is of shape (4, 4), followed by convolutions with (3, 3) kernels. Before the final fully-connected classification layer, we position the global average pooling (GAP) layer in order to accommodate the requirements of the CAM attribution technique.

\textbf{Hyperparameters.}
We trained the networks using Stochastic Gradient Descent (SGD). We explored various combinations of learning rate (LR), momentum, weight decay, and batch size. The range of LRs we evaluated was $10^{-6} - 10^0$. The tested batch sizes were $64$ and $500$.
The hyperparameters we picked for training were: LR =$10^{-4}$, momentum $=0.1$, weight decay $=0.1$, batch size $=64$.
The networks were trained for up to 500 epochs, with early stopping allowed after 10 consecutive epochs of validation loss rise, with warm-up of 50 epochs.

\section{Gradient-based interpretability techniques are fragile}
\label{app:CAM_is_fragile}
\textbf{Introduction to class activation mapping.} 
CAM is an attribution technique tailored for use with CNNs and leverages their design. A CNN produces different representations of the input data during a forward pass through the network and encodes them in different channels. CAM relies on the latent representation of the data present in the $N_{\text{ch}}$ output channels of the last convolutional layer in a network (activations $A_k$), by performing their weighted sum with weights $\alpha_k$ \cite{jung2021better},
\begin{equation}
    \text{CAM}^C(A) = \text{ReLU}~\qty(\sum_{k=1}^{N_{\text{ch}}} \alpha_k A_k)\,.
\end{equation}
\begin{figure}[t]
\centering
\includegraphics[width=0.95\columnwidth]{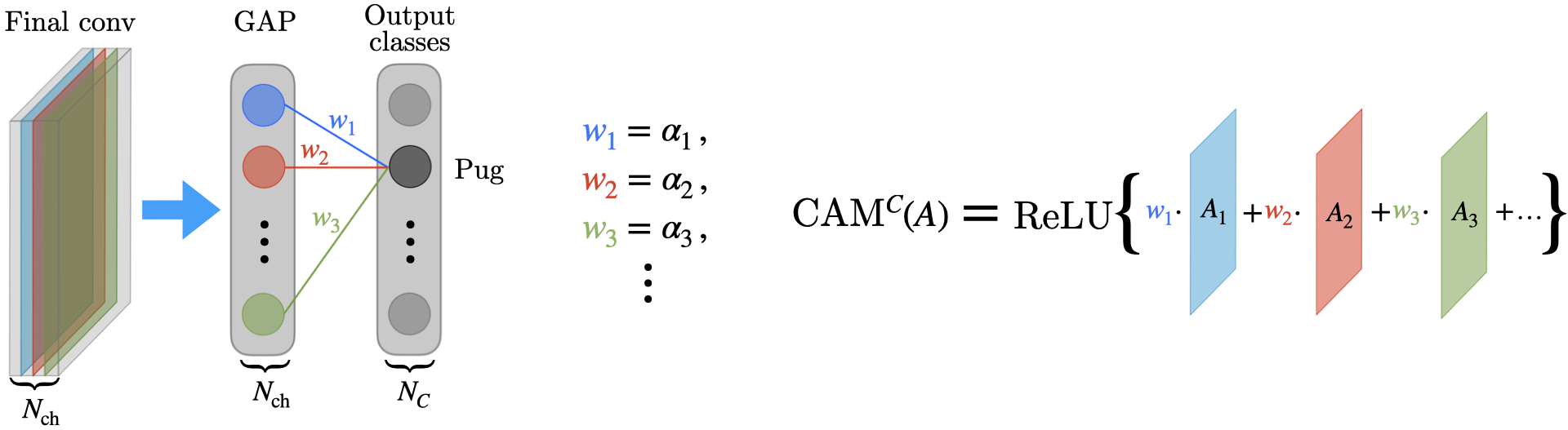}
\caption{\textbf{Class activation mapping (CAM)} generates a map that highlights relevant parts of the input data for predicting a selected class $C$ (out of $N_{C}$ classes), e.g., a pug out of all 1000 ImageNet categories. In the final convolutional layer of the network, each of the $N_{ch}$ channels generates a convolved image, $A_k$ (three out of $N_{ch}$ are marked with blue, red, and green squares in the figure). Each activation can encode some data features relevant to some of the classes. An intuitive example of such features could be the shortened snout of a pug or cat's whiskers. The GAP layer then reduces each of these activations to its mean value (e.g., each colored square got reduced to the scalar of the same color). Finally, the fully-connected layer maps GAP values to the network outputs and its weights are used as weights for the CAM. Adapted from \reflab~\cite{Zhou16CAM}.}
\label{fig:cam_graphic}
\end{figure}
The resulting map, after rescaling back to the original data size, highlights the areas that were the most influential in predicting a considered class $C$. The crucial part is the choice of the weights $\alpha_k$. The original formulation of the method relies on reducing each channel output of the last convolutional layer to a single number with global average pooling (GAP) layer~\cite{Zhou16CAM}. The weights of fully-connected layer connecting those numbers with the output corresponding to the considered class $C$ are then used as $\alpha_k$ for the weighted average of convolved images. We present this explanation graphically in \fig~\ref{fig:cam_graphic}. 

\begin{figure}[t]
    \centering
    \includegraphics[width=\columnwidth]{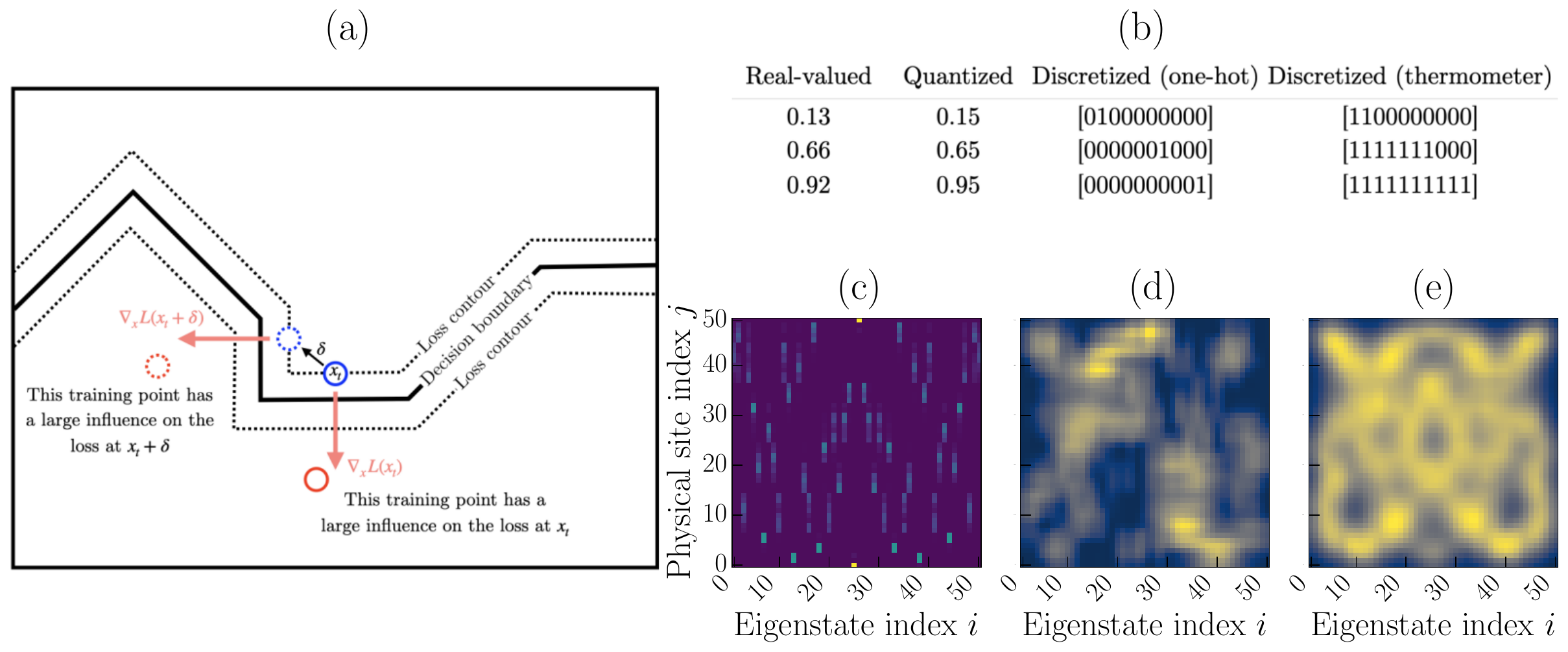}
    \caption{\textbf{Interpretation fragility.} (a) Gradient-based attribution techniques (CAM included) rely on the direction of the gradient (pink arrow), so small perturbation $\delta$ of test image $x_t$ can vary the basis of explanation, leaving class prediction unaffected. Panel adapted from \reflab~\cite{ghorbani_interpretation_2019}. (b) Thermometer encoding, a solution to this problem proposed by Buckman et al.~\cite{buckman_thermometer_2018}, is a discretization technique for encoding a floating point input variable as a vector representing its magnitude. Adapted from \reflab~\cite{buckman_thermometer_2018}. (d) In a disordered regime, CAM yields flawed interpretations. (c) The input is axially symmetric -- a property that the interpretation should preserve. (e) This is remedied by using thermometer encoding at the cost of limited network performance and a rise in computational cost. 
    }
    \label{fig:app:thermometer}
\end{figure}

\textbf{CAM fails for the disordered regime.} As discussed before, CAM provides invaluable insight into possible correlations in disorderless data. However, the results proved unreliable when applied to data from a disordered regime ($W / \inter \leq 0.05$). Contrary to the previously observed and expected tendencies, CAM importance maps were asymmetric and varied significantly between neighboring $\intra$ values and multiple realizations of the same point in phase space. This example is presented in \figs~\ref{fig:app:thermometer}(c)-(d).

\textbf{NN's fragility.} The CNNs have been shown to be prone to 'adversarial noise' attacks \cite{Yao2020, Duan21CVPR}. They can take various intricate forms, as discussed by \reflab~\cite{Wang23}. The simplest scenario of an adversarial noise attack is one where the input image is perturbed in a deliberately engineered way to fool the CNN and make it predict a different class while keeping the input visually indistinguishable for a human. This class of attacks targets the predictive power of CNN. However, the problem in our case is the more subtle -- predictions remain factual, and the gradient-based interpretation fails.

\textbf{Interpretation fragility.} The topic of attacks on interpretation alone has already been addressed by the computer science community \cite{ghorbani_interpretation_2019, dombrowski_explanations_2019, dombrowski_towards_2022}. The intuition provided by \reflab~\cite{ghorbani_interpretation_2019} is as follows. Gradient-based techniques, like CAM, rely on the direction one must go from any given test example to the decision boundary. This is the geometric interpretation of the gradient in parameter space. The decision boundary, being highly non-convex, makes it very easy to perturb a given test image while still within the boundaries of its class. This slight shift is enough for the gradient vector to point in another direction, producing a different interpretation. See \fig~\ref{fig:app:thermometer}(a).

\textbf{Thermometer encoding.} Buckman et al.~\cite{buckman_thermometer_2018} have proposed using thermometer encoding of input to remedy this problem. It belongs to a broader class of approaches to reducing problem complexity via casting input with floating point entries to discrete variables. A standard input discretization technique, one-hot encoding, encodes a quantized floating point number as a vector with the only non-zero entry marking the range in which it falls. \reflab~\cite{buckman_thermometer_2018} argues that by mimicking the behavior of a mercury thermometer, encoding the magnitude of the encoded input, the network gradients are better behaved. This change, they argue, reduces the fragility of gradient-based network attribution techniques. The thermometer encoding process is demonstrated in \fig~\ref{fig:app:thermometer}(b).

\textbf{Thermometer recovers CAM.} We have implemented thermometer encoding with the number of discretization bins ranging from 10 to 100 to test its applicability to our problem. We have successfully recovered the insight provided by CAM, as displayed in \fig~\ref{fig:app:thermometer}(e). We have found the $100$ discretization bins to be the optimal value for the problem. This technique retrieves CAM interpretation for all disorders in the range addressed by this study. 

\textbf{Cost of CAM interpretability.} While gradient-based interpretation is recovered, there is collateral damage done to network performance and training costs. None out of 100 networks trained using this thermometer encoding generalized well to data with disorder, even when trained on data with added small disorder $W / \inter \in \qty{0, 0.01, 0.05}$. This is a significant performance loss when compared to 22\% of well-generalizing networks obtained when the architecture is not adjusted to the thermometer encoding. There is also a substantial rise in the computational cost of network training. For $100$ discretization bins, the number of trainable parameters rises from $\mathcal{O}(10^3)$ to $\mathcal{O}(10^9)$. This, in turn, extended the training time of a single network from  $\mathcal{O}(1 \mathrm{min})$ to $\mathcal{O}(2 \mathrm{h})$. The training was performed in PyTorch \cite{paszke_pytorch_2019} on a CUDA-enabled NVIDIA GeForce GTX 1050 graphics card with 4\,GB\,of\,GDDR5\,VRAM.

\section{Toy model for CAM performance assessment}
\label{app:toy-cam}

\begin{figure}[t]
    \centering
    \includegraphics[width=0.75\columnwidth]{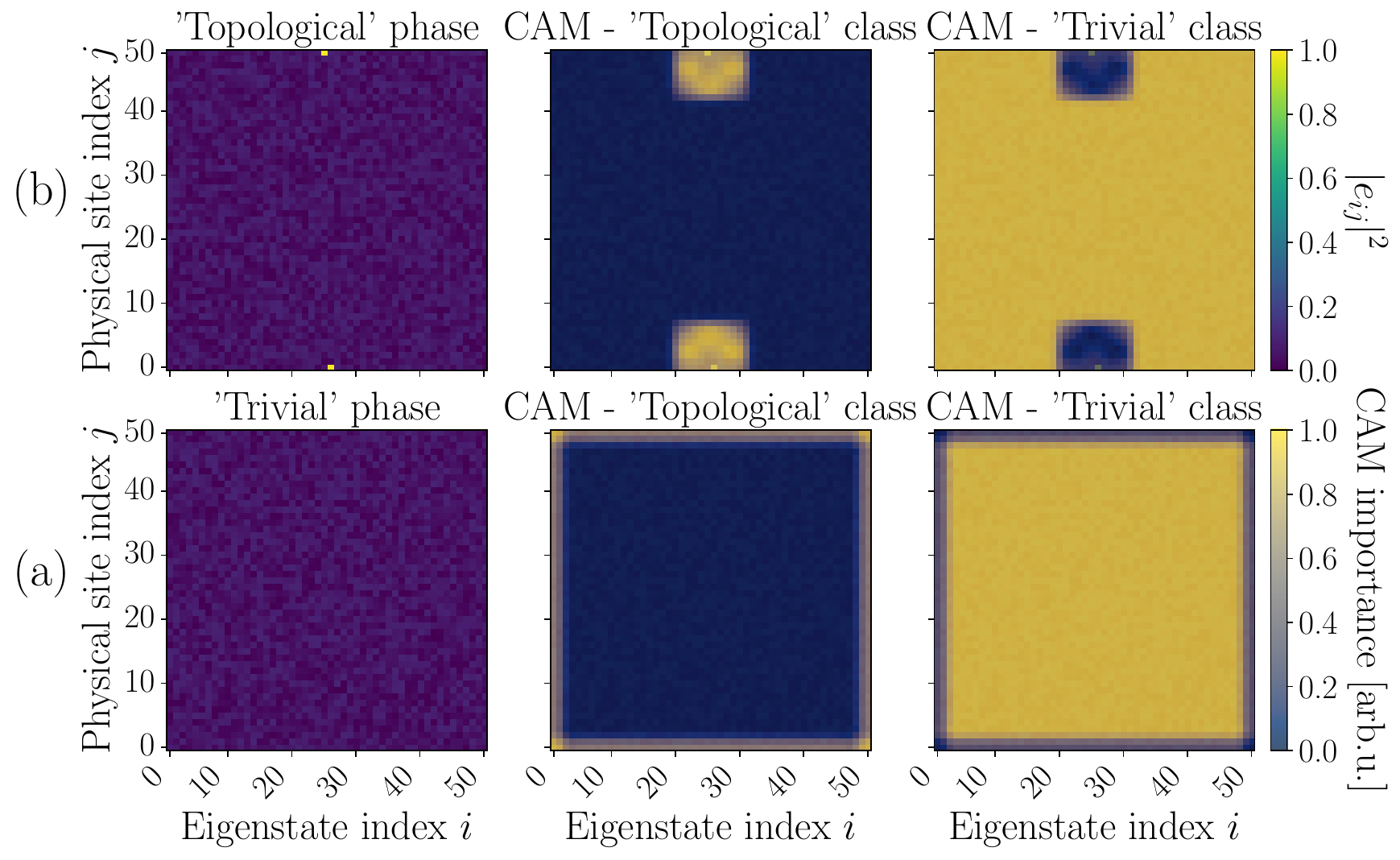}
    \caption{\textbf{CAM analysis of a network trained and tested in a simplified setting.} All entries in data points are drawn from the uniform distribution $U[0, 0.1]$. In half of the data, we set the edge entries of two middle columns to~1. This constitutes a 'Topological' phase, with the second part labeled 'Trivial' phase. CAM analysis of this setting confirms that the features the network uses for class discrimination are the bright pixels.}
    \label{fig:toy_model}
\end{figure}

The motivation behind our use of CAM can be demonstrated using a toy model of the problem at hand, imitating the SSH model with a slight disorder. In this setting, we create an artificial data set consisting of data points, where all entries are drawn from the uniform distribution $U[0, 0.1]$. We then separate it into two parts and set the edge entries of two middle columns to~1. 
In this setting, the phases are not distinguished by any correlations other than the two bright pixels we set. 
CAM analysis of a network trained and tested in this setting shows that the features the network uses for class discrimination are the bright pixels. This result is shown in \fig~\ref{fig:toy_model}.

\section{Evolution of the data representation across layers of a neural network}
\label{app:evolution_across_layers}
\textbf{Representation varies between network layers.}
We can also study the evolution of the data representation across layers, which gives additional insights into the behavior of different networks. To this end, we now investigate outputs of the three consecutive convolutional layers (Conv1, Conv2, Conv3), next to the output of the GAP layer, which we have studied in the main text in \fig~\ref{fig:clustering_final_layer}. Contrary to the discussion in the main text, here we do not remove the data from the vicinity of the transition point. We also inspect the respective representation of the test data without disorder and OOD test data with a single realization of multiple disorder amplitudes ($W / \inter \in \qty{0.15, 0.5, 1, 2}$). However, not all data sets can be visualized accurately in a few dimensions using only linear transformations. Therefore, we chose a nonlinear technique to visualize the underlying structure of the latent space representations in the CNNs more accurately than PCA, described in \seclab~\ref{sec:PCA_UMAP}. To this end, we use the uniform manifold approximation and projection (UMAP) \cite{UMAP18}.

\begin{figure}[b]
\begin{center}
\includegraphics[width=0.95\columnwidth]{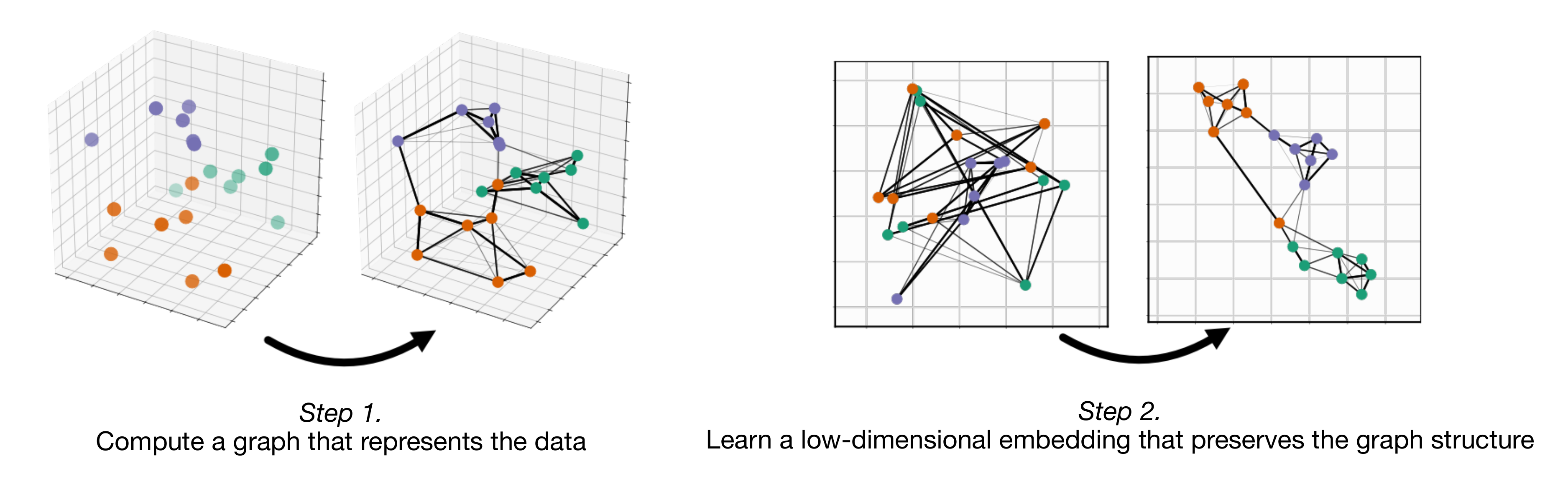}
\end{center}
\caption{\textbf{Uniform manifold approximation and projection (UMAP) routine sketch.} UMAP is a two-step routine: first, it computes a graph that represents the data. The second step is to learn a low-dimensional representation of the graph while trying to preserve the local, global, and topological structure of the data. This allows for efficient visualization of high-dimensional datasets. Adapted from \reflab~\cite{UMAP_docs}.}
\label{fig:app:UMAP}
\end{figure}

\begin{figure}[t]
    \centering
    \includegraphics[width=0.95\columnwidth]{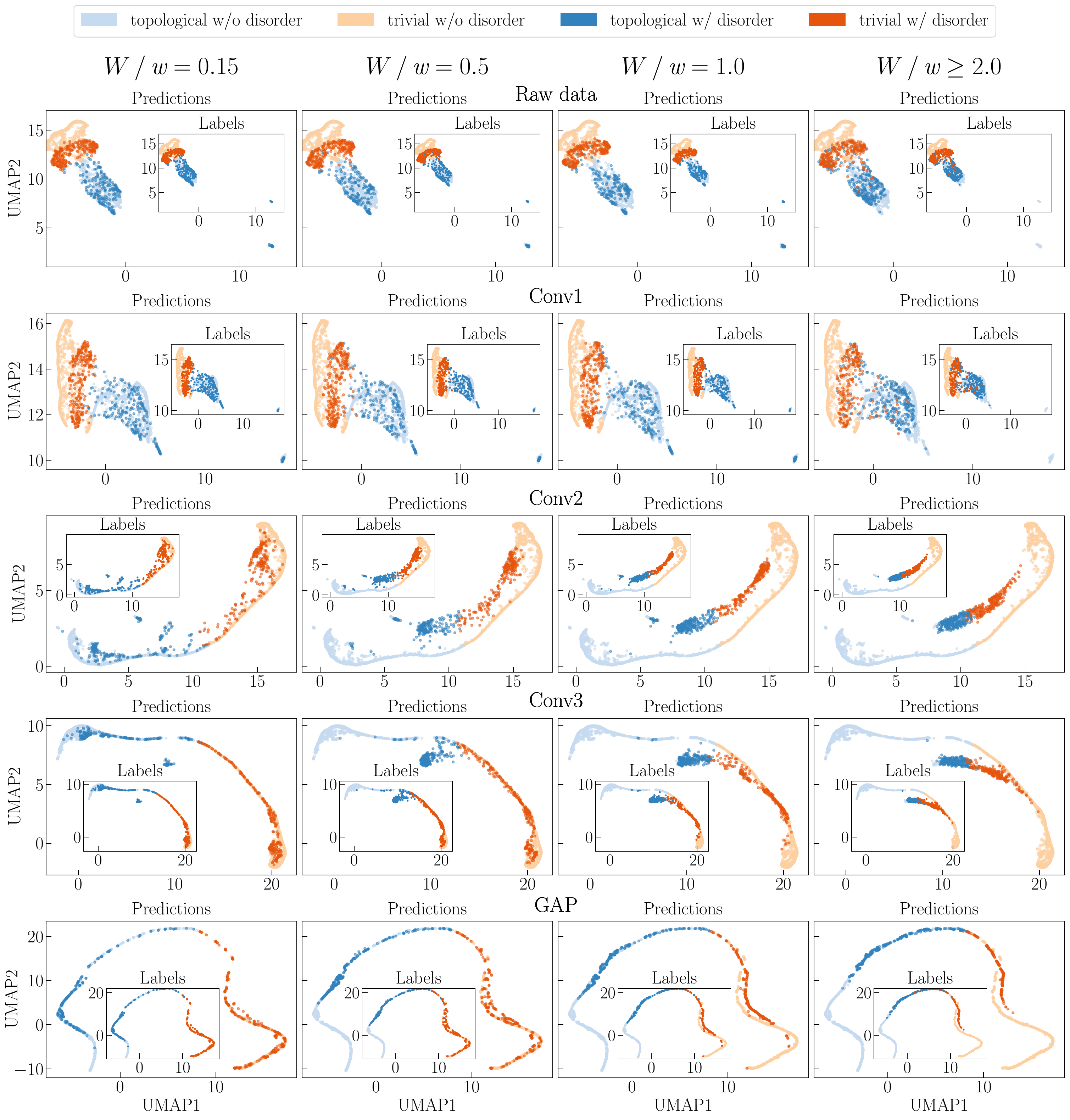}
    \caption{\textbf{Evolution of the internal data representation with UMAP within different layers of the well-generalizing CNN} that does not focus on edge states, as seen \fig~\ref{fig:cam_and_generalizations}(d),(h).}
    \label{fig:app:clustering_across_layers_smart}
\end{figure}

\begin{figure}[t]
    \centering
    \includegraphics[width=0.95\columnwidth]{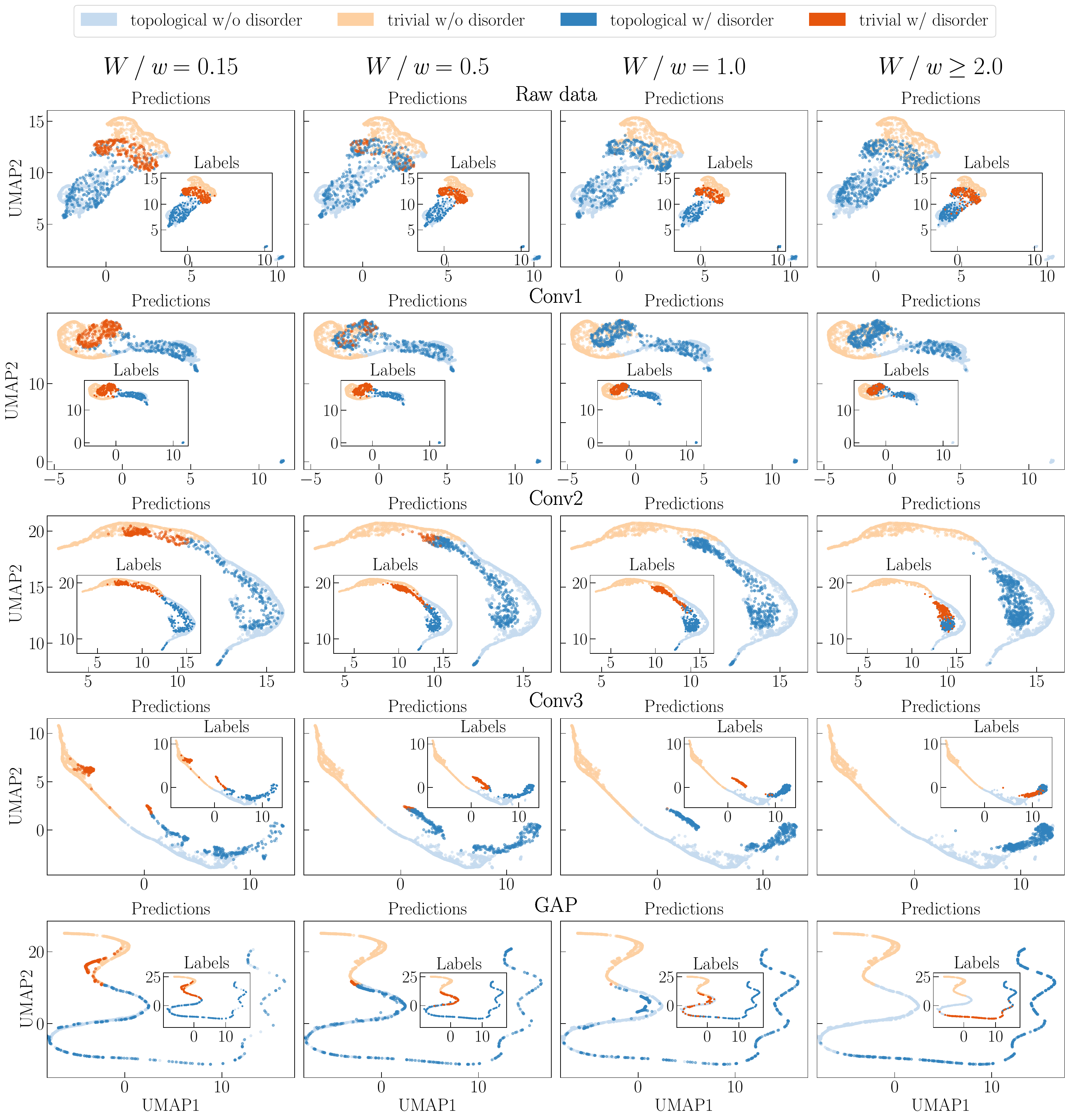}
    \caption{\textbf{Evolution of the internal data representation with UMAP within different layers of the poorly-generalizing CNN} that focuses on edge states, as seen \fig~\ref{fig:cam_and_generalizations}(c),(g).}
    \label{fig:app:clustering_across_layers_dumb}
\end{figure}

\textbf{UMAP -- nonlinear dimensionality reduction.} 
As seen in \fig~\ref{fig:app:UMAP}, UMAP is a two-step routine: first, it computes a graph that represents the data and then learns a low-dimensional representation of the graph while trying to preserve the local, global, and topological structure of the data. UMAP is good at visualizing high-dimensional nonlinear data sets, but it has some drawbacks. First, the embedding learning process is stochastic, which means that the reproducibility of the results is limited. Second, it can distort the true nature of data connectivity and density \cite{Cooley2019, wattenberg2016how, Chari2023}, so it is essential to be cautious when using it to conclude the clustering or concentration of data. It is a good practice to verify that the resulting number of clusters agrees with what we expect to find in the data based on results from other techniques. Despite its limitations, UMAP has proven to be computationally inexpensive and is often used because of its ability to visualize high-dimensional data efficiently. These are the reasons behind our decision to use it to analyze the latent space representations in the CNNs we studied. The UMAP embeddings are visualized in \figs~\ref{fig:app:clustering_across_layers_smart} and \ref{fig:app:clustering_across_layers_dumb} for well- and poorly-generalizing networks, respectively.

\textbf{UMAP results for raw data.}
Analysis of raw data with UMAP (first row of \figs~\ref{fig:app:clustering_across_layers_smart} and \ref{fig:app:clustering_across_layers_dumb}) allows for two interesting observations. Firstly, surprisingly, the UMAP embedding is itself a better classifier of data from disordered regimes than the majority of CNNs. The embedding of both classes in such data overlaps with the embeddings of classes from a disorderless system. This could allow for an informed decision about proper labels in the OOD regime. 
Secondly, it is a testament to the previously mentioned stochasticity of low-dimensional representations. The raw data subject to analysis are the same in both figures, but their embedding is slightly different. This is a result of UMAP's limited reproducibility.
Upon careful inspection, these UMAP embeddings led to one more observation: that the data coming from the topological class formed two disjoint clusters. This phenomenon is due to the hybridization of the edge states \cite{asboth_short_2016}.

\textbf{Well-generalizing CNNs represent data with and without disorder similarly.}
Let us first focus on the two-dimensional data representation generated by UMAP for the well-generalizing network, as seen in \fig~\ref{fig:app:clustering_across_layers_smart}. 
In the consecutive columns, we plot the disorderless data and the data with increasing disorder strength (indicated with darker orange and blue). In the first convolutional layer (Conv1), the disordered data are represented very similarly to those without the disorder. In other words, their representation overlaps. In the following two layers (Conv2, Conv3), the class representations overlap for intermediate disorder amplitudes ($W / \inter \in \qty{0.5, 1}$. With increasing disorder amplitude ($W / \inter \geq 2$), the embedded data start to disconnect from the original clusters. In particular, the embedded Conv3 layer activations, which initially clustered with the topological class, begin forming a class of its own. This may signify a change in the data structure the network has learned to observe. It also means that at this stage of the forward pass, the network has a disjointed representation only of strongly disordered data.

The lowest row presents embedded activations from the GAP layer, a qualitatively different picture. Knowing the nature of GAP output, it seems safe to assume that what we see here is an embedding of a unidimensional manifold. As disorder increases, the initially disjoint clusters of disorderless data gradually connect with data from increasingly disordered regimes. This captures well the perceived similarity of both classes an observer, artificial or human, has when observing the input data.

\textbf{Poorly-generalizing CNNs have very different representation of data with and without disorder.}
Let us now move on to the two-dimensional data representation generated by UMAP for the poorly-generalizing network, as seen in \fig~\ref{fig:app:clustering_across_layers_dumb}. 
In the first convolutional layer (Conv1), the disordered data are represented very similarly to those without the disorder. This behaviour is consistent with the one observed for the well-generalizing network.
In the following two layers (Conv2, Conv3), the class representations overlap for intermediate disorder amplitudes ($W / \inter \in \qty{0.5, 1}$. With increasing disorder amplitude ($W / \inter \geq 2$), the embedded data start to disconnect from the original clusters. Given the poor generalizability we expect from this network, it is not surprising to see them merge together, and then finally be positioned near just one of the initial distributions. The fact that it gets positioned near the topological phase representation might suggest that this would be the phase predicted in high disorder regime, and this notion turns out to be true here.

The lowest row of \fig~\ref{fig:app:clustering_across_layers_smart} is once again a qualitatively different picture from the previous layers, but of the same nature as embedding of GAP layer activations from well-generalizing network. This further strengthens the assumption that is an embedding of a unidimensional manifold. Here, however, it is hard to say surely that the initial clusters of disorderless data are well separated. Their, already low, separation vanishes as disorder increases. The disordered data then proceed to merge together and position in the range of topological class. This, once again, correctly suggests that this would be the phase predicted in high disorder regime.

\bibliography{literature}

\end{document}